\documentclass[aps,prx,twocolumn,showpacs,superscriptaddress,amsmath,amssymb]{revtex4-1}
\usepackage{graphicx}
\usepackage{latexsym}
\usepackage{amsmath,amssymb}
\usepackage{bm} 
\usepackage{color}
\usepackage{epsfig}

\usepackage{graphicx}
\usepackage[colorlinks=true,linkcolor=blue,citecolor=blue,urlcolor=blue]{hyperref}
\usepackage{bbold}
\usepackage{gensymb}

\usepackage{pgfplots,pgfplotstable}
\tikzset{every axis plot post/.append style={solid, thin},every mark/.append style={solid,scale=1}}
\usetikzlibrary{decorations.markings}
\tikzset{mdar/.style ={decoration={markings,mark={at position 0.5 with {\fill (2pt,0)--(-2pt,2pt)--(-2pt,-2pt)--cycle;}}},postaction={decorate}}}
\tikzstyle{dsh}=[dash pattern=on 2.5pt off 1.5pt]
\tikzset{intcur/.style ={decorate,decoration={snake,amplitude=.5mm,segment length=3mm}}}
\tikzstyle{zba}=[dash pattern=on 1pt off 1pt,line width=2pt]

\usepackage{simplewick}

\definecolor{myblue}{rgb}{.93, .93, 1}

\AtBeginDocument{%
    \newwrite\bibnotes
    \def\bibnotesext{Notes.bib}
    \immediate\openout\bibnotes=\jobname\bibnotesext
    \immediate\write\bibnotes{@CONTROL{REVTEX41Control}}
    \immediate\write\bibnotes{@CONTROL{%
    apsrev41Control,author="08",editor="1",pages="0",title="0",year="1"}}
     \if@filesw
     \immediate\write\@auxout{\string\citation{apsrev41Control}}%
    \fi
}%

\def \a {\alpha}
\def \b {\beta}
\def \d {\delta}
\def \D {\Delta}

\def \ve {\varepsilon}

\def \g {\gamma}
\def \G{\Gamma}

\def \o {\omega}

\def \dag {\dagger}

\def \eqv {\equiv}
\def \apx {\approx}

\def \til {\tilde}

\def \dag {\dagger}

\newcommand{\intv}[1]{\int_{\mbf #1}}

\def \rar {\rightarrow}

\def \la {\langle}
\def \ra {\rangle}
\def \fr {\frac}
\def \lf {\left}
\def \ri {\right}

\newcommand{\epvl}[1]{\la#1\ra}

\def \Tr {\mathrm{Tr}}
\def \bece {\begin{center}}
\def \ence {\end{center}}
\def \beeq {\begin{equation}}
\def \eneq {\end{equation}}
\def \beal {\begin{aligned}}
\def \enal {\end{aligned}}
\def \bega {\begin{gathered}}
\def \enga {\end{gathered}}
\def \benu {\begin{enumerate}}
\def \ennu {\end{enumerate}}
\def \beit {\begin{itemize}}
\def \enit {\end{itemize}}
\def \bede {\begin{description}}
\def \ende {\end{description}}
\def \betb {\begin{tabular}}
\def \entb {\end{tabular}}
\def \bear {\begin{array}}
\def \enar {\end{array}}

\def \mbf {\mathbf}
\def \mbb {\mathbb}

\def \mca {\mathcal}

\def \bsb{\boldsymbol}
\def \txt {\text}

\newcommand{\comment}[1]{}

\begin{document}


\title{Complex charge density waves at Van Hove singularity on hexagonal lattices: Haldane-model phase diagram and potential realization in kagome metals $\text{AV}_3\text{Sb}_5$}

\author{Yu-Ping Lin}
\affiliation{Department of Physics, University of Colorado, Boulder, Colorado 80309, USA}
\author{Rahul M. Nandkishore}
\affiliation{Department of Physics, University of Colorado, Boulder, Colorado 80309, USA}
\affiliation{Center for Theory of Quantum Matter, University of Colorado, Boulder, Colorado 80309, USA}

\date{\today}

\begin{abstract}
We investigate how the real and imaginary charge density waves interplay at the Van Hove singularity on the hexagonal lattices. A phenomenological analysis indicates the formation of $3Q$ complex orders at all three nesting momenta. Under a total phase condition, unequal phases at the three momenta break the rotation symmetry generally. The $3Q$ complex orders constitute a rich Haldane-model phase diagram. When effective time-reversal symmetries arise under 1-site translations, the Dirac semimetals are protected. The breakdown of these symmetries gaps the Dirac points and leads to the trivial and Chern insulator phases. These phases are deformations of purely real and imaginary orders, which exhibit trivial site and/or bond density and chiral flux orders, respectively. The exotic single-Dirac-point semimetals also appear along the gapless phase boundary. We further show that the theoretical model offers transparent interpretations of experimental observations in the kagome metals $\text{AV}_3\text{Sb}_5$ with $\text{A}=\text{K},\text{Rb},\text{Cs}$. The topological charge density waves may be identified with the complex orders in the Chern insulator phase. Meanwhile, the lower-temperature symmetry-breaking phenomena may be interpreted as the secondary orders from the complex order ground states. Our work sheds light on the nature of the topological charge density waves in the kagome metals $\text{AV}_3\text{Sb}_5$ and may offer useful indications to the experimentally observed charge orders in the future experiments.
\end{abstract}

\maketitle

\section{Introduction}

The studies of Fermi liquid instabilities on the hexagonal lattices has received enormous interest in the past decade. While most of the interest is devoted to the graphene with honeycomb lattice \cite{gonzalez08prb,nandkishore12np,nandkishore12prl,wang12prb,kiesl12prb,jiang14prx,ntc}, the materials with triangular \cite{martin08prl,akagi10jpsj,akagi12prl,tieleman13prl,ntc,maharaj13prb} and kagome lattices \cite{yu12prb,kiesel13prl,wang13prb} have also been studied extensively. A particularly interesting setup for such analyses is the doping to the Van Hove singularity \cite{vanhove53pr}. At this doping, the fermiology of the three lattices become identical, with the caveat that the translation from lattice scale interactions to interaction constants in momentum space is nontrivial on the kagome lattice \cite{kiesel13prl, wang13prb}. The density of states is logarithmically divergent at the $M$-point saddle points of dispersion energy, leading to the amplification of correlation effects. These saddle points define a hexagonal Fermi surface with parallel edges, which further supports the Fermi surface nesting at three finite momenta. The combination of these two singular structures can trigger various types of Fermi liquid instabilities. It has been shown that the $d\pm id$ chiral superconductivity is the universal leading weak-coupling instability at the Van Hove doping for repulsive interactions on triangular or honeycomb lattices \cite{nandkishore12np}. The spin density waves can also arise away from the Van Hove doping, where the orders develop at all three nesting momenta. These ground states are known as the $3Q$ states, which can realize the chiral noncoplanar Chern insulator \cite{martin08prl} and the uniaxial half metal \cite{nandkishore12prl}. On the other hand, it was shown that the charge density waves may develop from the sublattice interference on the kagome lattice \cite{kiesel13prl,wang13prb}. The $M$-point charge density waves with unconventional features have also been studied in the transition metal dichalcogenides \cite{mcmillan75prb,ishioka10prl,vanwezel11epl}. More recently, the doping of graphene with intercalation shows a flattening of dispersion energy at the Van Hove doping \cite{mcchesney10prl}, leading to the high-order Van Hove singularity with power-law divergent density of states \cite{yuan19nc}. This turns the phase diagram into the competition between the $d\pm id$ chiral superconductivity and the ferromagnetism \cite{gonzalez13prb,classen20prb,lin20prb}.

While most of the works at the Van Hove singularity have focused on the real orders in the particle-hole channels, the imaginary orders have not received as much investigation. The imaginary particle-hole orders at finite momenta can realize staggered/loop currents on the lattice, which corresponds to the formation of intrinsic staggered fluxes \cite{affleck88prb,nayak00prb,chakravarty01prb}. Such flux orders may break the time-reversal symmetry spontaneously. The development of orders at all three nesting momenta can further trigger nontrivial band topology in the ground states. For the imaginary charge density waves, a Chern insulator can develop from the $3Q$ chiral flux order \cite{venderbos16prbcdw,lin19prb}. Meanwhile, a quantum spin Hall insulator can arise from the $3Q$ uniaxial spin flux order, which is a combination of two opposite chiral flux orders at opposite spins \cite{venderbos16prbsdw}. Whether these topological states can arise as the leading instability at the Van Hove doping becomes an interesting topic to explore. It has been shown that the imaginary charge density wave is degenerate with the real spin density wave with spin flavors $N_f=2$, and is further dominant universally at larger number of flavors $N_f\geq4$ \cite{lin19prb,classen19prb}. Meanwhile, the staggered currents have also been proposed in a $\pi$-flux triangular lattice \cite{tieleman13prl}, as well as in the doped chiral spin liquid \cite{song21prb}. These observations indicate the possibility of realizing the topological imaginary orders in the systems with hexagonal lattices. A Ginzburg-Landau analysis has been conducted to investigate how the according $d$-wave order interplays with the real orders \cite{maharaj13prb}. However, the $d$-wave order has been treated as a secondary order to the real site density order, and its purely imaginary structure has not been appreciated. A complete analysis where the real and imaginary orders are treated on equal footing is urged for the advanced understanding of complex charge density waves. Such an analysis may uncover unconventional phenomena where the topological imaginary orders contribute.

\begin{figure}[t]
\centering
\includegraphics[scale = 0.6]{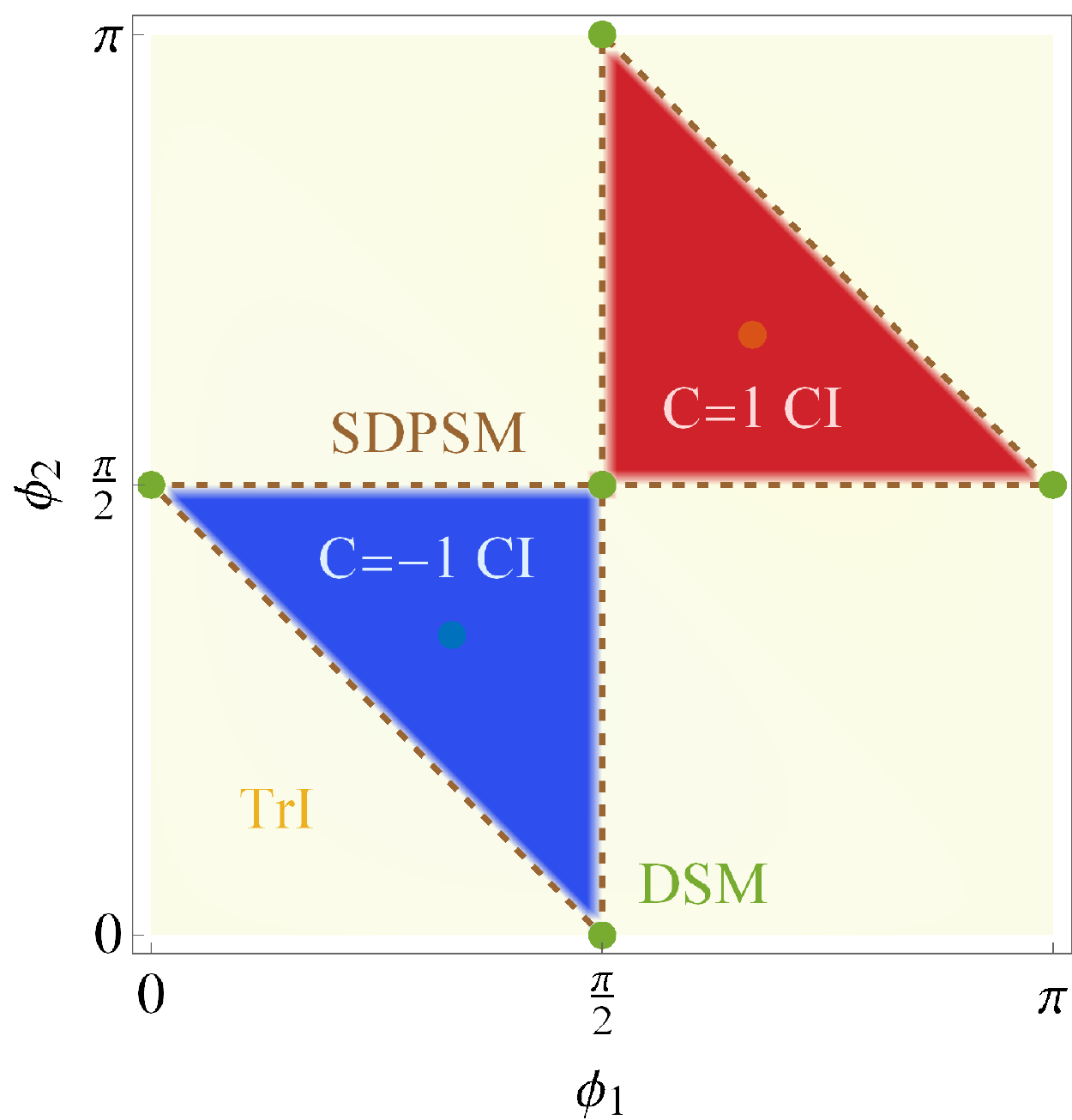}
\caption{\label{fig:pd} Haldane-model phase diagram of $3Q$ complex charge density waves from the computation of Chern number, where the $s$-wave order is chosen for the real order. Here the trivial insulator (TrI), Chern insulators (CI) with nonzero Chern numbers $C=\pm1$, Dirac semimetal (DSM), and single-Dirac-point semimetal (SDPSM) arise in different regimes. The minima of real order strength $\d_R^2$ under the total phase condition occur in the Chern insulator phases, below which the system evolves to the $3Q$ imaginary orders. Note that the phase diagram is periodic under $\phi_i\rar\phi_i+\pi$ for $i=1,2$.}
\end{figure}

Recently, a set of experiments observe unconventional $3Q$ charge density waves in the kagome metals $\txt{AV}_3\txt{Sb}_5$ with $\txt{A}=\txt{K},\txt{Rb},\txt{Cs}$ \cite{ortiz19prm,yang20sa,ortiz20prl,kenney21jpcm,jiang20ax,yu21prb,zhao21ax,liang21ax,uykur21ax,chen21axpdw,li21ax}, which occur at $80\txt{--}110\txt{ K}$ far above the superconductivity at $0.9\txt{--}2.7\txt{ K}$ \cite{ortiz21prm,zhao21axsc,chen21axpdw,chen21prl,duan21ax,zhang21prb}.
These orders develop at all three $M$ points and manifest giant anomalous Hall effects \cite{yang20sa,yu21prb}. Furthermore, a more exotic $1Q$ charge density wave is observed at a half $M$ point at lower temperature, which is accompanied by another rotation symmetry breaking effect along the same direction \cite{zhao21ax,chen21axpdw}. The signal of pair density wave at the three-quarter $M$ point is also observed along the same direction \cite{chen21axpdw}. A recent experiment indicates that the charge density waves may arise from the electronic repulsion instead of the strong electron-phonon coupling \cite{li21ax}. Given the proximity of the Fermi surface to the Van Hove singularity, the theoretically proposed chiral flux order \cite{venderbos16prbcdw,lin19prb} may contribute significantly to the topological response in these kagome metals.

In this work, we investigate how the real and imaginary charge density waves interplay at the Van Hove singularity on the hexagonal lattices. A phenomenological analysis indicates the formation of $3Q$ complex orders at all three nesting momenta. Under a total phase condition, unequal phases at the three momenta break the rotation symmetry generally. The $3Q$ complex orders constitute a rich Haldane-model phase diagram (Fig.~\ref{fig:pd}). When effective time-reversal symmetries arise under 1-site translations, the Dirac semimetals are protected. The breakdown of these symmetries gaps the Dirac points and leads to the trivial and Chern insulator phases. These phases are deformations of purely real and imaginary orders, which exhibit trivial site and/or bond density and chiral flux orders, respectively. The exotic single-Dirac-point semimetals also appear along the gapless phase boundary. We further show that the theoretical model offers transparent interpretations of experimental observations in the kagome metals $\text{AV}_3\text{Sb}_5$. The topological charge density waves may be identified with the complex orders in the Chern insulator phase. Meanwhile, the lower-temperature symmetry-breaking phenomena may be interpreted as the secondary orders from the complex order ground states. Our work sheds light on the nature of the topological charge density waves in the kagome metals $\text{AV}_3\text{Sb}_5$ and may offer useful indications to the experimentally observed charge orders in the future experiments.

\section{Van Hove fermiology on hexagonal lattices}

\begin{figure}[b]
\centering
\includegraphics[scale = 1]{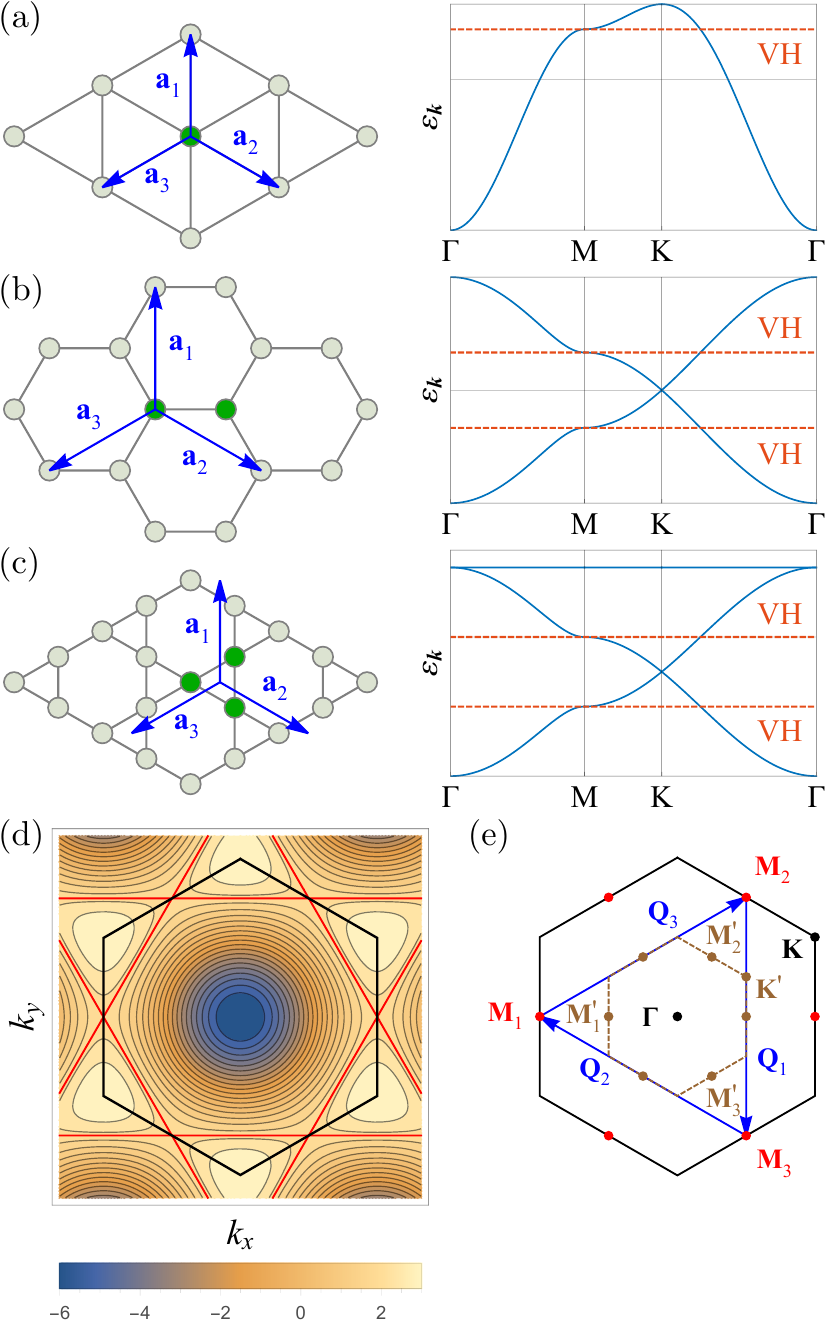}
\caption{\label{fig:lattices} The hexagonal lattices and their Van Hove (VH) fermiology. The (left) lattices and (right) nearest-neighbor tight-binding band structures are presented for (a) triangular, (b) honeycomb, and (c) kagome lattices. The origins of triangular Bravais lattices are defined by the intersections of lattice vectors $\mbf a_{\a=1,2,3}$. (d) The contour illustration of band structure on the triangular lattice. Without loss of generality, we present the momentum-space computation only for the triangular lattice in this work. The Fermi surface (red) is a hexagon in the hexagonal Brillouin zone (black), where the corner saddle points sit at the zone edge centers. The opposite sides are parallel, leading to the Fermi surface nesting at three momenta $\mbf Q_\a$. (e) Patch model of the Van Hove fermiology. The Fermi surface is approximated by the patches at the three inequivalent saddle points $\mbf M_\a$, which are connected by the nesting momenta $\mbf Q_\a$. The inner hexagon indicates the reduced Brillouin zone under the charge density waves.}
\end{figure}

We consider the fermionic models doped to the Van Hove singularity on the hexagonal lattices. These include the triangular, honeycomb, and kagome lattices [Figs.~\ref{fig:lattices}(a)--\ref{fig:lattices}(c)], all of which exhibit the same Van Hove fermiology in the hexagonal Brillouin zone [Fig.~\ref{fig:lattices}(d)]. For the triangular lattice with a single band, the Van Hove singularity occurs at the $3/4$ doping. For later convenience, we interpret this band as a hole band from the full doping, where the Van Hove singularity sits at the $-1/4$ doping. The honeycomb and kagome lattices contain two and three bands, respectively, where a pair of bands are separated by the Dirac points with opposite relative energies. In these systems, the Van Hove singularity occurs at the $\pm1/4$ dopings on the particle and hole bands, respectively. The Van Hove singularity is carried by the saddle points of dispersion energy, where the density of states becomes logarithmically divergent. For the hexagonal lattices, these saddle points sit at the three inequivalent zone edge centers $\mbf M_{\a=1,2,3}$. The Fermi surface takes these saddle points as the corners and form a hexagon in the Brillouin zone. Since the opposite Fermi lines are parallel to each other with opposite energy structures, a strong Fermi surface nesting is manifest at the Van Hove singularity. Note that the nesting vectors $\mbf Q_\a\eqv\mbf M_\a$ are half of reciprocal lattice vectors. This allows the Umklapp scattering to occur, from which various Fermi liquid instabilities may be triggered.

Due to the Van Hove singularity, the three saddle points dominate the rest parts of the Fermi surface at low energy. The low-energy effective theory is well described by the patch model \cite{nandkishore12np,lin19prb}, where the Fermi surface is approximated by three patches at these saddle points [Fig.~\ref{fig:lattices}(e)]
\beeq
H^0=\sum_{\a=0}^3(\ve_\a-\mu)\psi_\a^\dag\psi_\a.
\eneq
Here $\psi_\a$ with $\a=1,2,3$ are the fermions in the three patches with dispersion energy $\ve_\a$, and the chemical potential $\mu=0$ is defined at the Van Hove singularity. We have included a patch $\a=0$ at the zone center $\bsb\G$, which is coupled to the saddle points by the nesting momenta. Despite the distance from the Fermi surface, it may still contribute by lifting the degeneracy between otherwise degenerate orders. Note that the fermion flavor is suppressed since our analysis focuses on the charge orders.

\section{Phenomenology of complex charge density waves}

The combination of Van Hove singularity and Fermi surface nesting leads to the $\ln^2(1/T)$ divergences in the temperature $T$. These divergences can induce various Fermi liquid instabilities in the presence of interactions. Our interest lies in the charge density waves, which are the particle-hole condensates at finite momenta. At the Van Hove singularity on the hexagonal lattices, these condensates develop at the three nesting momenta $\mbf Q_\a$
\beeq
\D_{\a,\mbf k}=\epvl{\psi_{\mbf k+\mbf Q_\a}^\dag\psi_{\mbf k}}.
\eneq
The charge density waves are generally complex. While the real orders manifest the periodic modulations of charge site and/or bond densities, the imaginary order hosts the staggered currents. We will study the interplay between these orders in the framework of Ginzburg-Landau theory and determine the phase diagram of complex charge density waves.

\subsection{Irreducible pairing channels}

\begin{figure}[t]
\centering
\includegraphics[scale = 1]{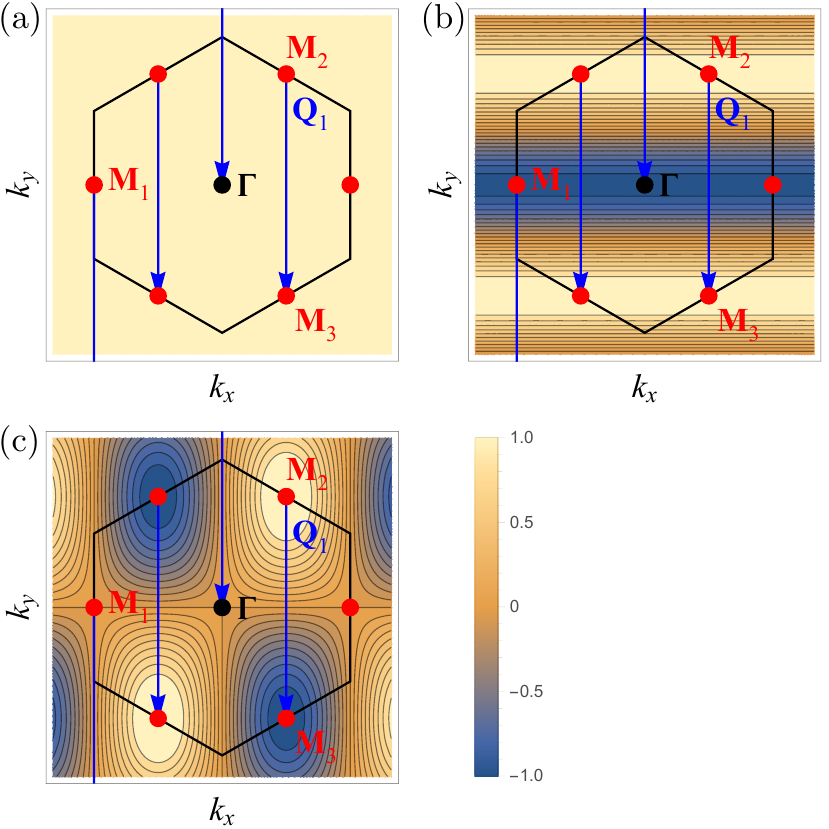}
\caption{\label{fig:dw} The form factors of the charge density waves with (a) real $s$-wave $f_{\a,\mbf k}^{s}$, (b) real $d_R$-wave $f_{\a,\mbf k}^{d_R}$, and (c) imaginary $d_I$-wave $f_{\a,\mbf k}^{d_I}$ orders in the momentum space. The momentum $\mbf Q_1$ and the connected pairs of points $(\mbf M_3,\mbf M_2)$, $(\mbf M_1,\bsb\G)$ are indicated in the figures. The form factors at the other two momenta $\a=2,3$ can be obtained by $\txt{C}_3$ rotations.}
\end{figure}

The irreducible pairing channels of charge density waves can be determined based on the symmetry and the momenta \cite{venderbos16prbcdw}. An important feature of charge density waves at $\mbf M_\a$ is the commensuration of momenta $\mbf Q_\a$. This results in the decoupling of real and imaginary orders into different irreducible pairing channels. The decoupling can be identified from the order function $\D_{\a,\mbf k}=\D_\a f_{\a,\mbf k}$. Here $\D_\a\in\mbb C$ is the order parameter and the form factor $f_{\a,\mbf k}\in\mbb R$ is the eigenmode of momentum shift $\til{\txt{T}}_{\mbf Q_\a}$. Under the commensuration, the condition $\til{\txt{T}}_{\mbf Q_\a}^2=\til{\txt{T}}_{2\mbf Q_\a}=1$ implies the eigenvalues $\pm1$ for the eigenmodes $f_{\a,\mbf k+\mbf Q_\a}=\til{\txt{T}}_{\mbf Q_\a}f_{\a,\mbf k}=\pm f_{\a,\mbf k}$. Meanwhile, a complex conjugate constraint is imposed on the order function $\D_{\a,\mbf k+\mbf Q_\a}=\epvl{\psi_{\mbf k+2\mbf Q_\a}^\dag\psi_{\mbf k+\mbf Q_\a}}=\D_{\a,\mbf k}^*$. Combining these two conditions, the order parameter acquires the purely real or imaginary form $\D_\a=\pm\D_\a^*$. Such a decoupling has also been confirmed from the flows of the renormalization group \cite{lin19prb}.

We adopt particular real and imaginary irreducible pairing channels under the commensurate conditions \cite{venderbos16prbcdw}. In the $s$- and $d_{R,I}$-wave channels, the explicit form factors read (Fig.~\ref{fig:dw})
\beeq\beal
f_{\a,\mbf k}^{s}&=1,\quad
f_{\a,\mbf k}^{d_R}=-\cos(\mbf k\cdot\mbf a_\a),\\
f_{\a,\mbf k}^{d_I}&=\cos(\mbf k\cdot\mbf a_\b)-\cos(\mbf k\cdot\mbf a_\g),\quad \g>\b>\a,
\enal\eneq
where $\mbf a_\a$ are the lattice vectors [Figs.~\ref{fig:lattices}(a)--\ref{fig:lattices}(c)]. We have defined the patch numbers $\a$ in a cyclic notation $1<2<3<1$. The $s$- and $d_R$-wave orders manifest the real condition $f^{s,d_R}_{\a,\mbf k+\mbf Q_\a}=f^{s,d_R}_{\a,\mbf k}$, thereby exhibiting the site and bond density modulations. Meanwhile, the $d_I$-wave order obeys the imaginary condition $f^{d_I}_{\a,\mbf k+\mbf Q_\a}=-f^{d_I}_{\a,\mbf k}$ and leads to staggered/loop currents. The form factors can be translated into the patch representations in the patch model. The real orders exhibit the $s$- or $d_R$-wave patch representation $(f_{\a,\mbf M_\a},f_{\a,\mbf M_\b},f_{\a,\mbf M_\g})=(\pm1,1,1)$ with $\g>\b>\a$, while the imaginary order carries the $d_I$-wave patch representation $(0,1,-1)$.

Our analysis focuses on the $s$- and $d_{R,I}$-wave irreducible pairing channels under the symmetry. However, the realistic structures of charge density waves may experience some deviations from these channels. Such deviations may be attributed to the strong suppression of condensates away from the Fermi surface, such as in the pairing between a saddle point $\mbf M_\a$ and the zone center $\bsb\G$ in the $s$- and $d_R$-wave real orders. The combination of different channels may resolve this issue. For example, the combined $(s+d_R)$-wave real order exhibits the patch representation $(0,1,1)$, which involves only the saddle points at the Van Hove singularity. This configuration is sufficient for the suppression effects in the patch model.

\subsection{Ginzburg-Landau free energy}

Having identified the real and imaginary charge density waves, we introduce the interactions in these two channels and obtain the interacting theory
\beeq
H=H^0+\fr{1}{2}\sum_{O=R,I}\sum_\a g^O(P^O_\a)^\dag P^O_\a.
\eneq
The pairing operators at $\mbf M_\a$
\beeq\beal
P^R_\a&=\lf.\txt{Re}[\psi_\g^\dag\psi_\b]\ri|_{\g>\b>\a}\pm\txt{Re}[\psi_\a^\dag\psi_0],\\
P^I_\a&=\lf.\txt{Im}[\psi_\g^\dag\psi_\b]\ri|_{\g>\b>\a}
\enal\eneq
are defined according to the patch representations. Note that the zone center $\bsb\G$ is coupled to the saddle points $\mbf M_\a$ only in the real channel. The $s$- and $d_R$-wave orders correspond to the $\pm$ signs, respectively.

We assume that both of the real and imaginary orders can develop below certain critical temperatures $T^{R,I}_c$, where the originally positive interactions become negative $g^{R,I}<0$. To study the interplay between these two orders, we conduct a coherent path integral and extract the mean-field free energy \cite{lin19prb}. Under a Hubbard-Stratonovich transformation, the interactions are decoupled by the bosonic complex order parameter $\vec\D=(\D_1,\D_2,\D_3)$. The real and imaginary components of the order parameter $\D_\a=\D^R_\a+i\D^I_\a=|\D_\a|e^{i\phi_\a}$ are coupled to the pairing operators $(P^{R/I}_\a)^\dag$, respectively. Integrating out the fermionic modes, we arrive at the mean-field free energy
\beeq
f=\fr{2}{|g^R|}|\vec\D^R|^2+\fr{2}{|g^I|}|\vec\D^I|^2-\Tr\ln(-\mca G^{-1}).
\eneq
Here the trace denotes the momentum-frequency summation $\Tr\sim T\sum_n\intv{k}$. The inverse Green's function takes the form
\beeq
\mca G^{-1}=\lf(\bear{cccc}G_1^{-1}&\D_3&\bar\D_2&\pm\D^R_1\\\bar\D_3&G_2^{-1}&\D_1&\pm\D^R_2\\\D_2&\bar\D_1&G_3^{-1}&\pm\D^R_3\\\pm\D^R_1&\pm\D^R_2&\pm\D^R_3&G_0^{-1}\enar\ri),
\eneq
where the free propagators are defined $G_\a=[i\o-(\ve_\a-\mu)]^{-1}$ with the fermionic Matsubara frequency $\o$.

We expand the free energy with respect to the infinitesimal order parameters near the critical temperature $T_c=\max\{T^R_c,T^I_c\}$. Ignoring the constant part, the expansion to the quartic order gives the Ginzburg-Landau free energy
\beeq\beal
f&=Z^{(2)}_R|\vec\D^R|^2+Z^{(2)}_I|\vec\D^I|^2\\&\hspace{11pt}-Z^{(3)}(\D_1\D_2\D_3+\bar\D_1\bar\D_2\bar\D_3)-6Z^{(3)}_0\D^R_1\D^R_2\D^R_3\\&\hspace{11pt}+\fr{1}{2}Z^{(4)}_1|\vec\D|^4+(Z^{(4)}_2-Z^{(4)}_1)\\&\hspace{11pt}\times(|\D_1|^2|\D_2|^2+|\D_2|^2|\D_3|^2+|\D_3|^2|\D_1|^2).
\enal\eneq
The quadratic prefactor $Z^{(2)}_I$ turns negative below $T^I_c$, indicating a second-order phase transition for the purely imaginary order. Meanwhile, the other quadratic prefactor $Z^{(2)}_R$ may remain infinitesimally positive at $T^R_c$ and turn negative at lower temperature, since a cubic term supports a first-order phase transition. The isotropic quartic prefactor $Z^{(4)}_1=\Tr(G_1^2G_2^2)=\Tr(G_2^2G_3^2)=\Tr(G_3^2G_1^2)>0$ remains positive and ensures the stability of Ginzburg-Landau free energy. The charge density waves develop below $T_c$ and expand a large order manifold, where the degeneracy is reduced by the cubic and quartic anisotropies. At the cubic order, the primary anisotropy reads $\sim2|\D_1||\D_2||\D_3|\cos(\phi_1+\phi_2+\phi_3)$ with prefactor $Z^{(3)}=\Tr(G_1G_2G_3)$. The magnitude part $|\D_1||\D_2||\D_3|$ indicates that the `$3Q$ orders' are energetically favored, where the orders at the three momenta develop simultaneously with the same magnitude
\beeq
|\D_1|=|\D_2|=|\D_3|.
\eneq
The same conclusion follows from the consideration of the quartic anisotropy with a negative prefactor $Z^{(4)}_2-Z^{(4)}_1<0$, where $Z^{(4)}_2=\Tr(G_1^2G_2G_3)=\Tr(G_2^2G_3G_1)=\Tr(G_3^2G_1G_2)>0$ \cite{nandkishore12prl}. The phase degeneracy of the $3Q$ orders is lifted by the phase part of cubic anisotropies. The primary cubic anisotropy $\sim\cos(\phi_1+\phi_2+\phi_3)$ imposes a total phase condition on the $3Q$ orders. Furthermore, the secondary cubic anisotropy with the prefactor $Z^{(3)}_0=\Tr(G_0G_1G_2)=\Tr(G_0G_2G_3)=\Tr(G_0G_3G_1)$ leads to the additional preference of real orders. Note that this term is secondary since it involves the zone center $\bsb\G$ away from the Fermi surface, and only comes into play if the real and imaginary orders would be degenerate in the absence of this term. 

The phase conditions from the cubic anisotropies deserve further discussions. For the hole bands with the zone center $\bsb\G$ at the band bottoms, the cubic prefactors are positive $-Z^{(3)},-6Z^{(3)}_0>0$. On the contrary, the prefactors are negative $-Z^{(3)},-6Z^{(3)}_0<0$ for the particle bands, where the zone center $\bsb\G$ lies at the band tops. According to the prefactors of the primary cubic anisotropy, we summarize the total phase condition as
\beeq
\phi_1+\phi_2+\phi_3=
\begin{cases}
(2n+1)\pi,&\txt{hole band}\\
2n\pi,&\txt{particle band}
\end{cases}
\eneq
with $n\in\mbb Z$. The preference of real orders from the secondary cubic anisotropy then follows directly. Notably, the total phase condition suggests that the phases at the three momenta are generally different. This indicates that rotation symmetry breaking occurs generally in the $3Q$ complex orders. In the real channel, the secondary cubic anisotropy suggests an energetically favored imbalance between the $s$- and $d_R$-wave orders. This is also confirmed by a Ginzburg-Landau analysis where both orders are included explicitly.

\subsection{Energetically favored ground states}

\begin{figure}[b]
\centering
\includegraphics[scale = 1]{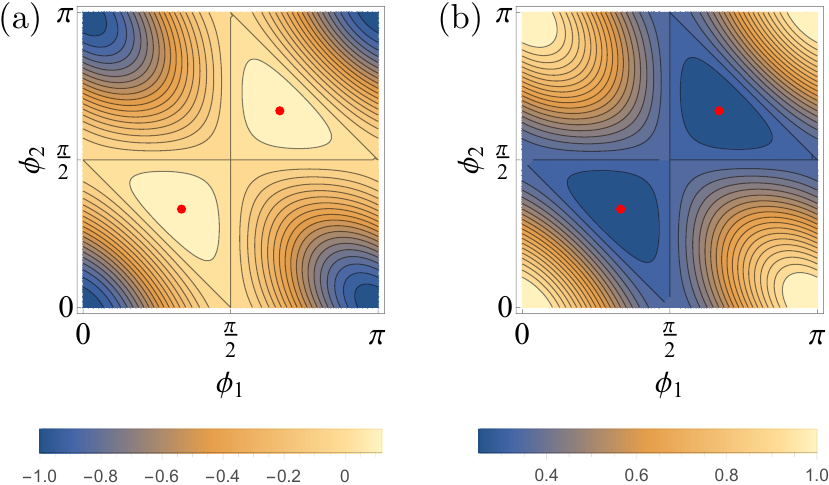}
\caption{\label{fig:cubic} (a) The secondary cubic anisotropy $\D^R_1\D^R_2\D^R_3\in[-1,1/8]$ and (b) the strength of real order $\d_R^2\in[1/4,1]$. Here two of the phases $\phi_{1,2}$ are tuned, while the third phase is given by $\phi_3=\pi-(\phi_1+\phi_2)$ under the total phase condition. The corners exhibit the minimal $\D^R_1\D^R_2\D^R_3=-1$ and the maximal $\d_R^2=1$. Meanwhile, the red points $(\pi/3,\pi/3,\pi/3)$ and $(2\pi/3,2\pi/3,-\pi/3)$ exhibit the maximal $\D^R_1\D^R_2\D^R_3=1/8$ and the minimal $\d_R^2=1/4$. Despite different functional forms, the contours in the two figures are identical. The states with the same $\d_R^2$ are degenerate under the secondary cubic anisotropy. In the balanced case, the system sits at the corners with minimal secondary cubic anisotropy and exhibit the real orders with maximal $\d_R^2=1$. As the upper bound $\bar\d_R^2$ decreases in the imbalanced case, the contour at $\d_R^2=\bar\d_R^2$ defines the degenerate ground states with the lowest available secondary cubic anisotropy.}
\end{figure}

The energetically favored ground states can now be determined from the Ginzburg-Landau free energy. We start by assuming that the real and imaginary orders are balanced, where equal strength of orders can be manifest near $T_c$. (We will shortly relax this assumption.) The full order manifold contains all $3Q$ orders for any strength of real order $\d_R=|\vec\D_R|/|\vec\D|\in[0,1]$. For example, the real and imaginary orders carry $\d_R=1$ and $0$, respectively, while $0<\d_R<1$ for the complex orders. The cubic anisotropies select the real orders as the energetically favored ground states (Fig.~\ref{fig:cubic}). For the hole bands, the negative order $(\phi_\a,\phi_\b,\phi_\g)=(\pi,\pi,\pi)$ with $\g>\b>\a$ and the 1-negative orders $(0,0,\pi)$ are energetically favored. Meanwhile, the particle bands prefer the positive order $(0,0,0)$ and the 2-negative orders $(0,\pi,\pi)$.

Things may become different when the real and imaginary orders are imbalanced. When the two orders develop at distinct critical temperatures $T^R_c\neq T^I_c$, their available magnitudes below the critical temperature $T_c$ become different. This constrains the available range of $\d_R$ and shrinks the order manifold, thereby changing the energetically favored ground states. A renormalization group study shows that the imaginary order is much stronger than the real order under the electronic repulsion \cite{lin19prb}. Nevertheless, the imaginary order may bring up the real order so as to minimize the free energy. We study the Ginzburg-Landau free energy in this complex-order regime (Fig.~\ref{fig:cubic}), where an upper bound $\bar\d_R\geq\d_R$ with $\bar\d_R<1$ indicates the deviation from the balanced case. As $\bar\d_R$ decreases, the energetically favored ground state is pushed away from the real orders. The degenerate ground states are determined by the energy contour of the secondary cubic anisotropy at this $\bar\d_R$. Note that $\bar\d_R$ reaches the minimum $\bar\d_R^\txt{min}=1/2$ at $(m_1\pi\pm\pi/3,m_2\pi\pm\pi/3,m_3\pi\pm\pi/3)$ with $m_{1,2,3}\in\mbb Z$ under the total phase condition. The total phase condition is broken below this point, and the system gradually evolves to the imaginary orders $(\pm\pi/2,\pm\pi/2,\pm\pi/2)$. On the other hand, the real order may be stronger than the imaginary order under the strong phonon-mediated attraction. The sublattice interference on the particle band of kagome lattice may also drive the system into this regime \cite{kiesel13prl,wang13prb}. According to the secondary cubic anisotropy, the real orders remain energetically favored under the imbalance.

\section{Ground state phase diagram}

\begin{figure}[b]
\centering
\includegraphics[scale = 1]{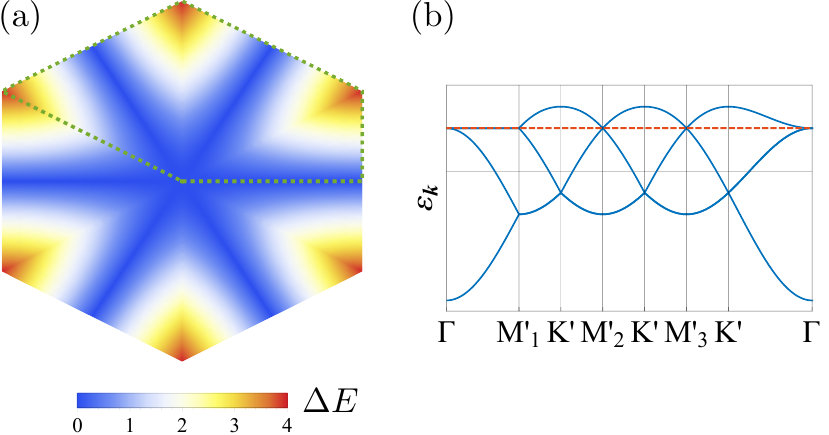}
\caption{\label{fig:rbz} Fermiology in the absence of charge density waves. (a) The band gap at the Fermi level and (b) the band structure along the indicated path (green dashed polygon) in the reduced Brillouin zone. The Fermi surface is composed of three lines connecting opposite $\pm\mbf M_\a'$. These lines cross at the zone center $\bsb\G$ and form a topological quadratic band crossing.}
\end{figure}

The Ginzburg-Landau analysis identifies various complex orders as the ground states of $3Q$ charge density waves. These orders may lead to unconventional phenomena, such as the Chern insulators with quantized intrinsic Hall conductivity \cite{venderbos16prbcdw,lin19prb}. To examine the ground state properties, we consider the mean-field model of $3Q$ complex orders and analyze the according band structures. Since the charge density waves break the 1-site translation symmetries, the system exhibits $2\times2$ enlarged unit cells. The nesting momenta $\mbf Q_\a$ serve as the new reciprocal lattice vectors, which connect the new reciprocal lattice sites $\mbf M_\a$ and $\bsb\G$. In the $1/2\times1/2$ reduced Brillouin zone [Fig.~\ref{fig:lattices}(e)], the mean-field Hamiltonian takes a four-band form
\beeq
\mca H_{\mbf k}=\lf(\bear{cccc}\ve_{\mbf M_1,\mbf k}&-\D_{3,\mbf M_1,\mbf k}&-\D_{2,\mbf M_1,\mbf k}&-\D_{1,\mbf M_1,\mbf k}\\-\D_{3,\mbf M_2,\mbf k}&\ve_{\mbf M_2,\mbf k}&-\D_{1,\mbf M_2,\mbf k}&-\D_{2,\mbf M_2,\mbf k}\\-\D_{2,\mbf M_3,\mbf k}&-\D_{1,\mbf M_3,\mbf k}&\ve_{\mbf M_3,\mbf k}&-\D_{3,\mbf M_3,\mbf k}\\-\D_{1,\mbf\G,\mbf k}&-\D_{2,\mbf\G,\mbf k}&-\D_{3,\mbf\G,\mbf k}&\ve_{\mbf\G,\mbf k}\enar\ri).
\eneq
The reciprocal lattice sites with dispersion energies $\ve_{\mbf P,\mbf k}=\ve_{\mbf k-\mbf P}$ are coupled by the charge density waves $\D_{\a,\mbf P,\mbf k}=\D_{\a,\mbf k-\mbf P}=|\D_\a|(\cos\phi_\a f_{\a,\mbf k-\mbf P}^{s,d_R}+i\sin\phi_\a f_{\a,\mbf k-\mbf P}^{d_I})$.

In the absence of charge density waves, the Fermi surface is composed of three straight lines connecting opposite zone edge centers $\pm\mbf M_\a'=\pm\mbf M_\a/2$ (Fig.~\ref{fig:rbz}). These Fermi lines cross at the zone center $\bsb\G$ and lead to a triply degenerate quadratic band crossing. Importantly, the quadratic band crossing inherits the $d$-wave structure of saddle points and manifest topological $\pm2\pi$ phase winding \cite{sun09prl,chern12prl}. Such a topological band crossing is protected by the time-reversal and $\txt{C}_6$ rotation symmetries. The charge density waves may break the symmetries and gap the topological band crossing \cite{chern12prl,venderbos16prbcdw,lin19prb}. The resulting bands can inherit the nontrivial $\pm2\pi$ phase winding, thereby forming topologically nontrivial states. On the other hand, the doubly degenerate band crossings at $\mbf M'_\a$ are protected by the symmetries of 1-site translations at $\mbf a_{\b\ne\a}$, inversion $\txt{C}_2$, and time reversal \cite{venderbos16prbcdw}. Although $\txt{C}_2$ symmetry remains present under the charge density waves, the band crossings can be gapped by the breakdown of 1-site translation and time-reversal symmetries. Fully gapped insulators can appear from the gapping of these band crossings.

In addition to the analysis of band structures, we also map out the site, bond, and current density modulations on the hexagonal lattices. The Fourier transform of charge density waves
\beeq\beal
&\epvl{\psi_{\mbf r}^\dag\psi_{\mbf r'}}\\&\hspace{11pt}=\sum_{\g>\b>\a}\lf\{\D_\a e^{i(-\mbf M_\g\cdot\mbf r+\mbf M_\b\cdot\mbf r')}+\bar\D_\a e^{i(-\mbf M_\b\cdot\mbf r+\mbf M_\g\cdot\mbf r')}\ri.\\&\hspace{22pt}\lf.\pm\D_\a^R\lf[e^{i(-\mbf M_\a\cdot\mbf r+\bsb\G\cdot\mbf r')}+e^{i(-\bsb\G\cdot\mbf r+\mbf M_\a\cdot\mbf r')}\ri]\ri\}
\enal\eneq
gives the densities of site $\rho_{\mbf r}=\txt{Re}[\epvl{\psi_{\mbf r}^\dag\psi_{\mbf r}}]$, bond $\rho_{\mbf r\mbf r'}=\txt{Re}[\epvl{\psi_{\mbf r}^\dag\psi_{\mbf r'}}]$, and current $j_{\mbf r\mbf r'}=\txt{Im}[\epvl{\psi_{\mbf r}^\dag\psi_{\mbf r'}}]$. For the honeycomb and kagome lattices with multiple bands, the evaluation involves a projection from the band eigenstate to the sublattice sites. Importantly, there is an asymmetry between the hole and particle band eigenstates on the kagome lattice. Each saddle point receives the contributions from two sublattice sites on the hole band, while the remaining sublattice site is involved singly on the particle band \cite{kiesel12prb}. The real-space pattern can indicate the residual symmetries under the charge density waves, which serves as an important complement to the determination of band properties.

We now examine whether the nontrivial band topology occurs in the potential ground states, thereby uncovering the phase diagram of $3Q$ complex charge density waves.

\begin{figure*}[t]
\centering
\includegraphics[scale = 1]{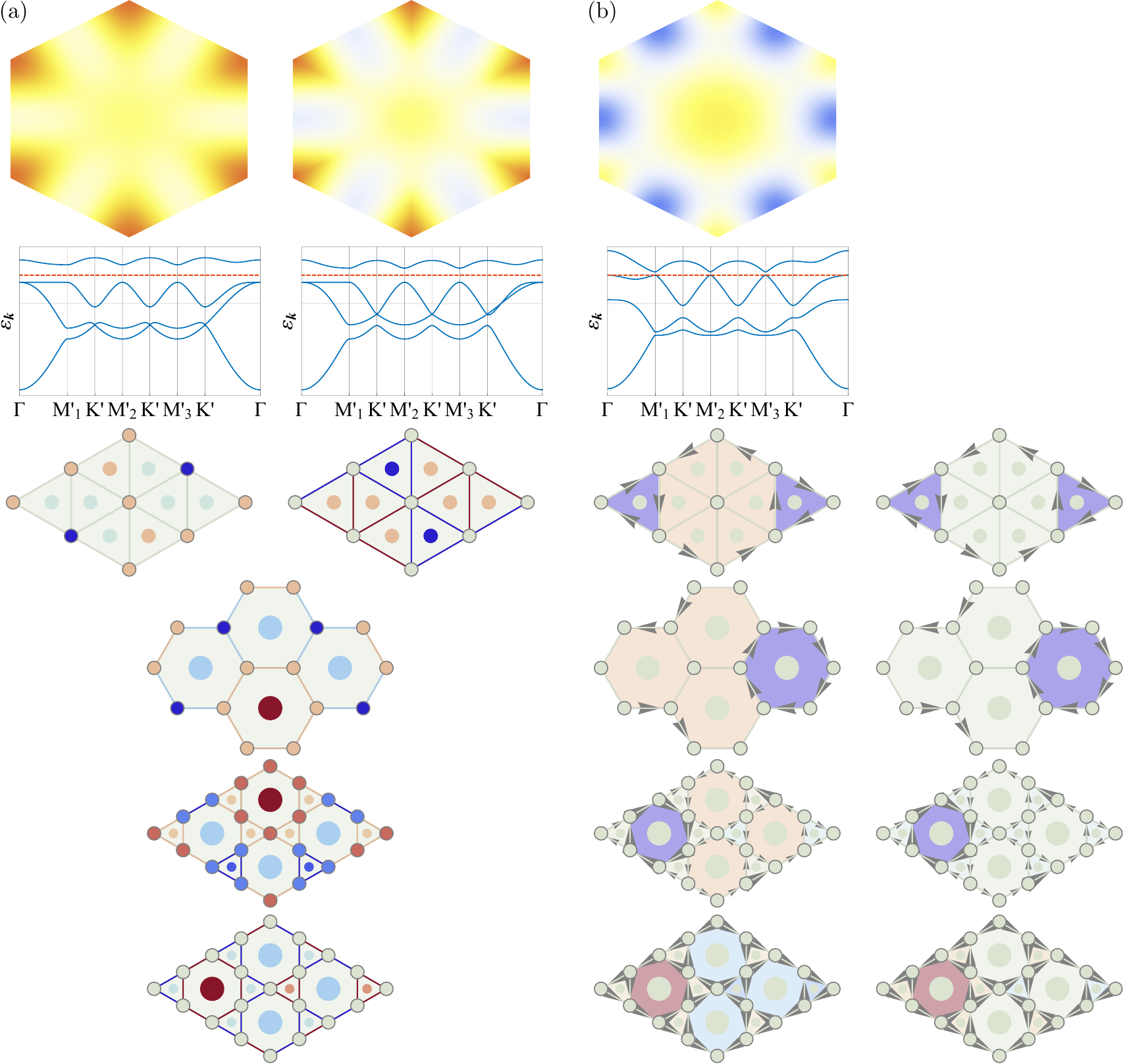}
\caption{\label{fig:trci} The (a) trivial and (b) Chern insulator ground states from the $3Q$ real $(0,0,\pi)$ [$(0,0,0)$ for the particle band on the kagome lattice] and imaginary $(\pi/2,\pi/2,\pi/2)$ charge density waves, respectively. The first row illustrates the band gaps at the Fermi level in the mean-field Hamiltonian, where the colorbar is the same as in Fig.~\ref{fig:rbz}(a). The second row presents the according band structures. The rest of the rows illustrate the patterns on the triangular and honeycomb lattices, as well as the kagome lattice at the (up) hole and (down) particle dopings. (a) For the trivial insulator, we present the (left) $s$- and (right) $d_R$-wave real orders on the triangular lattice, while the $(s+d_R)$-wave order is presented on the honeycomb and kagome lattices. (b) For the Chern insulator, we present the (left) staggered and (right) chiral flux orders. The site and bond densities are indicated by the colors, where the positive and negative values correspond to red and blue, respectively. The triangle and hexagon densities are summed over the surrounding sites and bonds and indicated by the center dots. The current strength is indicated by the arrow size. The flux in each triangle or hexagon is summed over the surrounding bond currents and indicated by the background.}
\end{figure*}

\subsection{Trivial insulator from real order}

When the orders at all three momenta are real [Fig.~\ref{fig:trci}(a)], the system preserves both the time-reversal and $\txt{C}_6$ rotation symmetries. This can be observed from the real-space patterns of density modulations, where a density order $(3,-1,-1,-1)$ is present in the $2\times2$ enlarged unit cell. For the triangular lattice, the $s$- and $d_R$-wave orders manifest the site and bond density modulations, respectively \cite{venderbos16prbcdw}. The realistic orders may be closer to the combined $(s+d_R)$-wave order, where the pairings between the saddle points $\epvl{\psi_\g^\dag\psi_\b}$ lead to the equally mixed site and bond density modulations. According to the secondary cubic anisotropy in the free energy, a secondary imbalance between these two orders is energetically favored. For the honeycomb lattice and the hole band on the kagome lattice, the $(s+d_R)$-wave order also shows an equally mixed site and bond density modulations from the pairings between the saddle points $\epvl{\psi_\g^\dag\psi_\b}$. Under the imbalance between the $s$- and $d_R$-wave orders, the finite condensates $\epvl{\psi_\a^\dag\psi_0}^{(*)}$ lead to secondary modulations. Similar results are observed on the particle band on the kagome lattice. However, the site density modulation is absent, which results from the single-sublattice-site structure of band eigenstate at the saddle points. Note that the bond density modulation shows the `inverse star-of-David' pattern.

With both time-reversal and $\txt{C}_6$ rotation symmetries, the topological quadratic band crossing remains stable at the zone center $\bsb\G$. However, the degenerate triplet is split into a singlet and a degenerate doublet, where the latter hosts the protected quadratic band crossing \cite{venderbos16prbcdw}. Whether the doublet sits at the Fermi level determines the gap opening. We find that the energetically favored ground states are those with the Fermi level lying between the doublet and the singlet. The gap is opened in the whole reduced Brillouin zone, leading to a trivial insulator with zero Chern number $C=0$. Note that the results are consistent with the maximization of ordering energy, which is equivalent to the maximization of gap structure. For the hole bands, the fully gapped states are the energetically favored ground states under the total phase condition $\phi_1+\phi_2+\phi_3=(2n+1)\pi$, such as $(0,0,\pi)$ and $(\pi,\pi,\pi)$. Meanwhile, the energetically unfavored states with $\phi_1+\phi_2+\phi_3=2n\pi$, including $(0,0,0)$ and $(0,\pi,\pi)$, exhibit a Fermi-level cut across the quadratic band crossing. The gap structure and energetic hierarchy are interchanged on the particle bands. The latter states become fully gapped and energetically favored, while the former states are gapless and energetically unfavored.

\subsection{Chern insulator from imaginary order}

\begin{figure}[t]
\centering
\includegraphics[scale = 1]{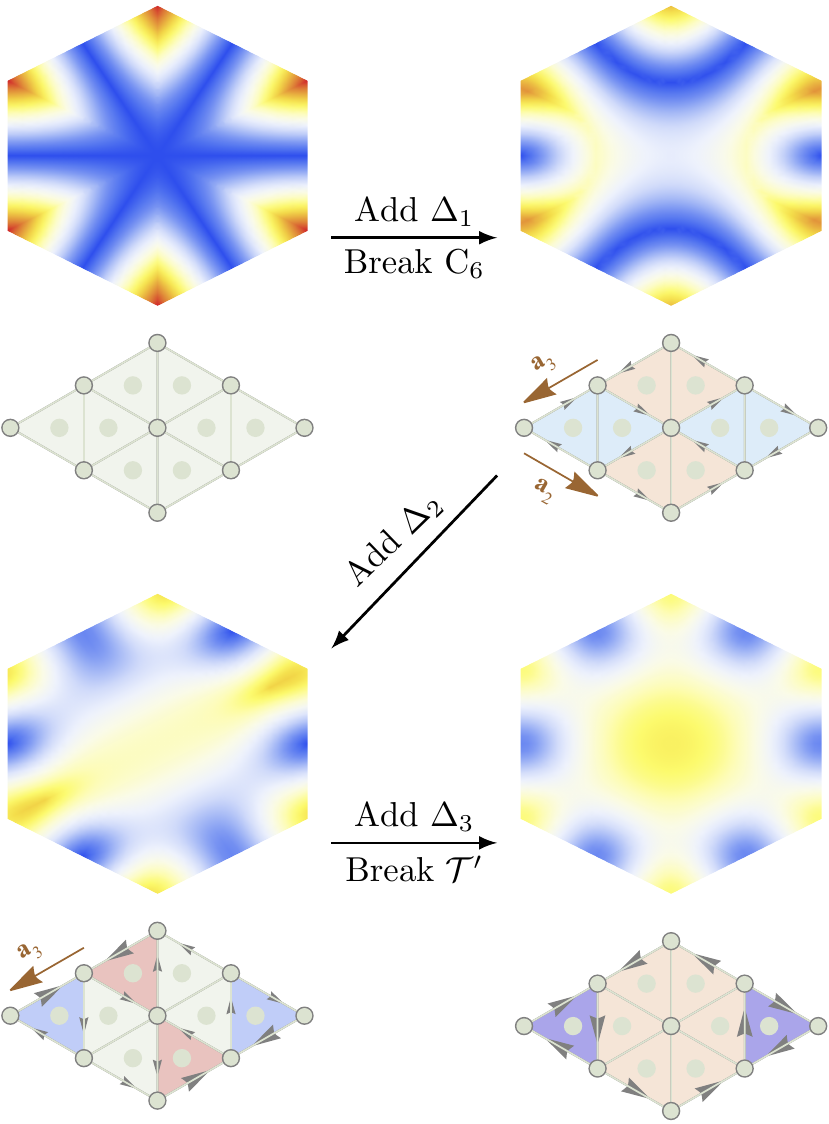}
\caption{\label{fig:cfo} The formation of Chern insulator can be understood from a step-by-step construction of $3Q$ imaginary order. The energy gap at the Fermi level in the reduced Brillouin zone and the real-space current pattern are illustrated in each step. The colorbar for the energy gap is the same as in Fig.~\ref{fig:rbz}(a). The side arrows in $1Q$ and $2Q$ orders indicate the available 1-site translations that support the effective time-reversal symmetries $\mca T'$.}
\end{figure}

For the charge density waves with imaginary orders at all three momenta $(\pm\pi/2,\pm\pi/2,\pm\pi/2)$ [Fig.~\ref{fig:trci}(b)], the staggered/loop currents are induced on the lattice. These currents are related to the intrinsic staggered fluxes in the triangles and hexagons, where a flux order $(3,-1,-1,-1)$ is manifest in the $2\times2$ enlarged unit cell \cite{venderbos16prbcdw,lin19prb}. While the $\txt{C}_6$ rotation symmetry is preserved, the time-reversal symmetry is broken spontaneously. This indicates the gap of topological quadratic band crossing at the zone center $\bsb\G$. The Fermi surface is fully gapped under the $3Q$ order, where the bands inherit the nontrivial $\pm2\pi$ phase winding of the quadratic band crossing. The ground state thus manifests a Chern insulator with nonzero Chern number $C=\pm1$ \cite{venderbos16prbcdw,lin19prb}.

\begin{figure*}[t]
\centering
\includegraphics[scale = 1]{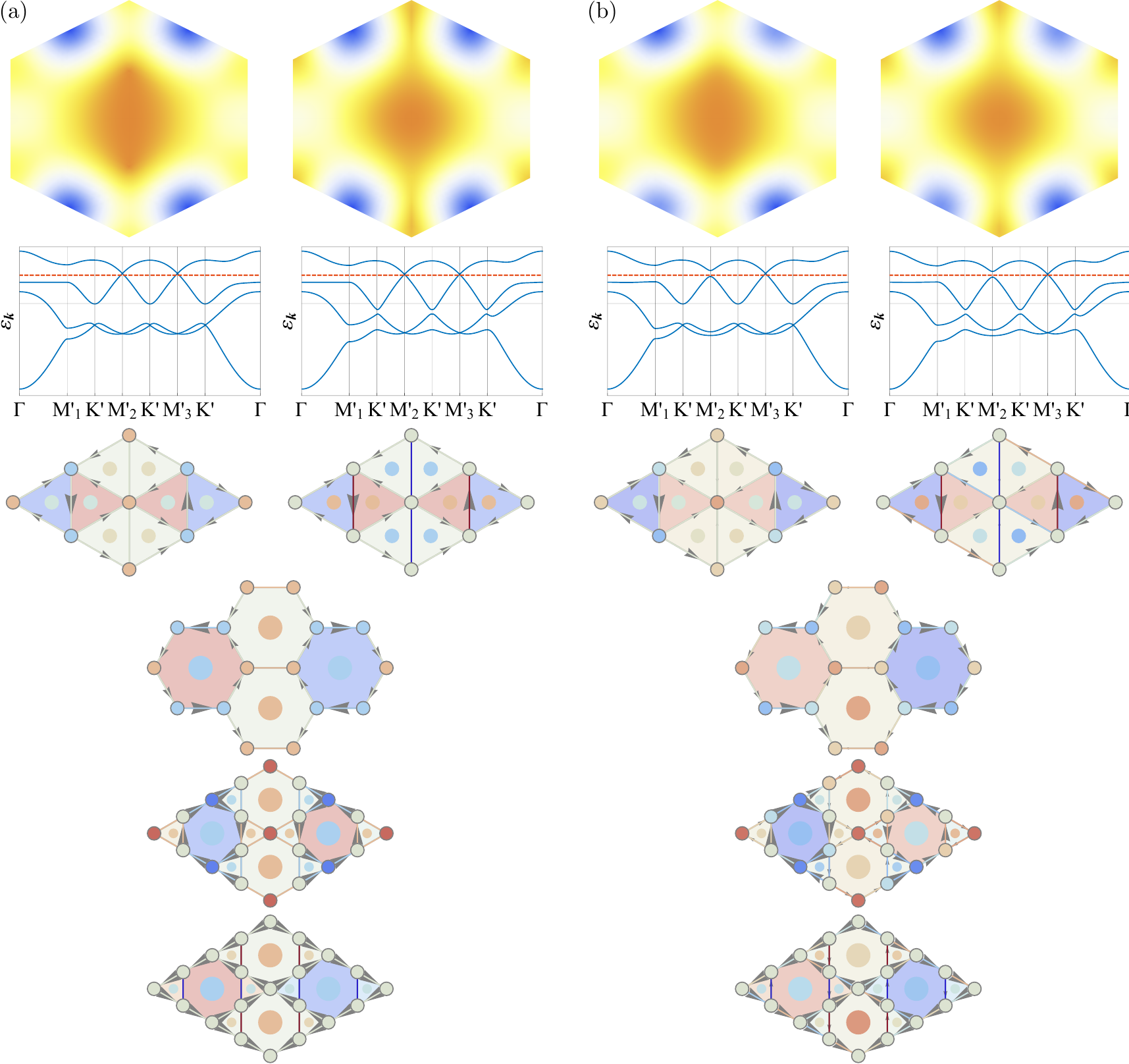}
\caption{\label{fig:sms} The semimetallic ground states of $3Q$ complex charge density waves. (a) Dirac semimetal from the complex orders at $(0,\pi/2,\pi/2)$ and $(\pi,\pi/2,\pi/2)$ on the hole and particle bands, respectively. (b) Single-Dirac-point semimetal from the complex order. When the real order is $s$-wave, we choose $(\pi[20/180],\pi[70/180],\pi[90/180])$ and $(\pi[200/180],\pi[70/180],\pi[90/180])$ on the hole and particle bands, respectively. In the case where the real order is $d_R$-wave, we choose $(\pi[20-\d/180],\pi[70/180],\pi[90+\d/180])$ with $\d=3.766$ on the triangular lattice. The patterns on the hole bands of the other lattices are also shown at this point, although the gapless points $\d$ are slightly different on different lattices. Similar setup is chosen for the particle band on the kagome lattice with $\d=8.6241$. The layout of the figures are the same as in Fig.~\ref{fig:trci}(a).}
\end{figure*}

To acquire more information on how the Chern insulator forms, we investigate how the gap structure evolves under the development of each order (Fig.~\ref{fig:cfo}) \cite{lin19prb}. The breakdown of symmetries plays an important role in this procedure. When the order develops at a single momentum $\mbf Q_\a$ and breaks the $\txt{C}_6$ rotation symmetry, the $1Q$ order gaps the topological quadratic band crossing at the zone center $\bsb\G$. The Fermi surface exhibits a Dirac point at $\mbf M_\a'$, as well as a curved nodal line connecting the other two $\mbf M_{\b\neq\a}'$. These nodal structures inherit the $\pm2\pi$ phase winding of the quadratic band crossing. Time-reversal symmetry is broken by the currents on the lattice. Nevertheless, effective time-reversal symmetries arise by combining 1-site translations at $\mbf a_{\b\neq\a}$. Each effective time-reversal symmetry at $\mbf a_{\b\neq\a}$ protects the band crossings at two edge centers $\mbf M'_{\g\neq\b}$. As a second order occurs at a momentum $\mbf Q_{\b\neq\a}$, the effective time-reversal symmetry at $\mbf a_{\g\neq\a,\b}$ is still present. The $2Q$ order gaps the nodal line into a Dirac point at $\mbf M'_\b$, leaving a pair of Dirac points at $\mbf M'_{\a,\b}$ on the Fermi surface. The third order at $\mbf Q_{\g\neq\a,\b}$ gaps the pair of Dirac points by breaking the effective time-reversal symmetry. Despite the restoration of $\txt{C}_6$ rotation symmetry, the topological quadratic band crossing at the zone center $\bsb\G$ remains gapped. The $3Q$ order thus turns the system into a Chern insulator with $C=\pm1$ as in the Haldane model \cite{haldane88prl}.

We note that the topological nature of the $3Q$ imaginary order may be interpreted more naturally with an alternative chiral flux order [Fig.~\ref{fig:trci}(b)]. The physical order in the imaginary order is the current order, which is unique for each choice of $3Q$ order. According to the pattern of current modulations, the flux order may be assigned as an auxiliary order in the state. The staggered flux orders are demanded in the $1Q$ and $2Q$ orders due to the effective time-reversal symmetry. Meanwhile, the absence of such a symmetry in the $3Q$ orders allows more freedom in choosing the flux orders. Although the $3Q$ imaginary orders inherit the staggered flux orders $(3,-1,-1,-1)$ from the $1Q$ orders, the chiral flux orders with intrinsic dilute fluxes $(1,0,0,0)$ may serve as more natural choices. These orders manifest an intrinsic flux in only one triangle or hexagon in each $2\times2$ enlarged unit cell. A unit anomalous Hall conductivity is induced accordingly, corresponding to the Chern number $C=\pm1$. The chiral flux orders also indicate the possibility of switching with external magnetic fields, which reflects the chiral nature of the states.

\subsection{Semimetals in-between}

We have identified the trivial and Chern insulators as the ground states of real and imaginary orders. A natural expectation is that the gapless states should also occur in the phase diagram, which serve as the critical states between the two gapped phases. Such gapless states may occur, for example, at $(m\pi,\pm\pi/2,\pm\pi/2)$ with $m\in\mbb Z$ [Fig.~\ref{fig:sms}(a)]. The real-space pattern indicates the presence of an effective time-reversal symmetry under a 1-site translation at $\mbf a_\a$, similar to the $2Q$ imaginary orders (Fig.~\ref{fig:cfo}). This protects a pair of Dirac points at $\mbf M'_{\b\neq\a}$ and leads to a Dirac semimetal. The effective time-reversal symmetry is broken away from this critical point $(m\pi+\d\phi,\pm\pi/2-\d\phi,\pm\pi/2)$ with $\d\phi\neq0$. Remarkably, one of the Dirac points becomes gapped, leaving only one Dirac point at the Fermi level [Fig.~\ref{fig:sms}(b)].  This exotic single-Dirac-point semimetal breaks the fermion doubling theorem, similar to the two-dimensional surfaces of three-dimensional topological insulators \cite{hasan10rmp}. The possibility of realizing a single Dirac point between the trivial and Chern insulators has already been anticipated in the Haldane model \cite{haldane88prl}. When an effective time-reversal symmetry occurs at $\d\phi=\pm\pi/2$, a new Dirac point appears and leads to a new Dirac semimetal. These semimetallic states constitute the gapless phase boundary between the trivial and Chern insulators in the phase diagram.

\subsection{Phase diagram}

Having analyzed the ground state properties at specific points, we now map out the phase diagram of the $3Q$ complex charge density waves. We determine the phase diagram under the total phase condition by computing the Chern number (Fig.~\ref{fig:pd}) \cite{fukui05jpsp}. When the real and imaginary orders are balanced at $\bar\d_R=1$, the ground states are the trivial insulators from the $3Q$ real orders $(m_1\pi,m_2\pi,m_3\pi)$ with $m_{1,2,3}\in\mbb Z$. As the system becomes imbalanced with decreasing $\bar\d_R<1$, the ground state remains in the trivial insulator phase until reaching the phase boundary. In the case where the real order is $s$-wave, the phase boundary line at $\bar\d_R^c=1/\sqrt3$ is defined by the phase condition $\phi_\a=\pm\pi/2$ for a single $\a$. When the real order is $d_R$-wave, the phase boundary becomes slightly curved. The Dirac semimetals develop at $(m\pi,\pm\pi/2,\pm\pi/2)$, where an effective time-reversal symmetry at $\mbf a_\a$ protects a pair of Dirac points at $\mbf M_{\b\neq\a}'$. The breakdown of this symmetry leads to the single-Dirac-point semimetals along the rest of the phase boundary. The further decrease of $\bar\d_R$ gaps the system into the Chern insulator with nonzero Chern number $C=\pm1$. The minima $\bar\d_R^\txt{min}=1/2$ under the total phase condition occur within the Chern insulator phases at $(m_1\pi\pm\pi/3,m_2\pi\pm\pi/3,m_3\pi\pm\pi/3)$ with $m_{1,2,3}\in\mbb Z$. As $\bar\d_R$ decreases further toward $\bar\d_R=0$, the ground state remains in the Chern insulator phase and evolves continuously to the $3Q$ imaginary orders $(\pm\pi/2,\pm\pi/2,\pm\pi/2)$.

The $3Q$ complex charge density waves constitute a phase diagram reminiscent of the one in the Haldane model \cite{haldane88prl,hasan10rmp,cooper19rmp}. A major difference lies in the manifestations of symmetries. In the Haldane model, the pair of Dirac points at $\pm\mbf K$ are related and protected by the inversion and time-reversal symmetries. When these symmetries are broken, the gaps of Dirac points lead to the trivial and Chern insulators. In the $3Q$ complex charge density waves, the pair of Dirac points at $\mbf M'_{\b\neq\a}$ are protected by the symmetries of inversion and effective time reversal at $\mbf a_\a$. However, the inversion symmetry now connects the equivalent points $\pm\mbf M_\a'$ and persists under the $3Q$ complex orders. The gaps of Dirac points are controlled only by the effective time-reversal symmetry breaking. Nevertheless, the resulting gapped states still manifest both the trivial and Chern insulators. Such a Haldane-model phase diagram may offer useful information to the experimentally observed charge orders on the hexagonal lattices.

\section{Topological charge density waves in kagome metals}

\begin{figure}[b]
\centering
\includegraphics[scale = 1]{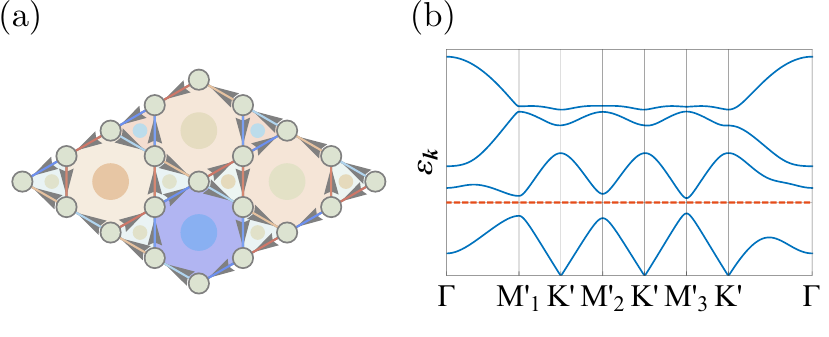}
\caption{\label{fig:exp} The proposed interpretation to the experimental results in the kagome metals $\txt{AV}_3\txt{Sb}_5$. (a) The real-space pattern of $3Q$ complex charge density wave at $(\pi[56/180],\pi[252/180],\pi[52/180])$. The bond density modulation shows a star-of-David pattern, and the according hexagon density pattern may be matched with the experimental result of $\txt{KV}_3\txt{Sb}_5$ \cite{jiang20ax}. (b) The band structure of complex charge density wave at $(\pi[60/180],\pi[231/180],\pi[69/180])$, which is assigned based on the experimental result of $\txt{CsV}_3\txt{Sb}_5$ \cite{zhao21ax}. We observe the hierarchy of anisotropic valley gaps $\D\ve_{\mbf M_3'}<\D\ve_{\mbf M_1'}<\D\ve_{\mbf M_2'}$ with similar differences $\D\ve_{\mbf M_1'}-\D\ve_{\mbf M_3'}\gtrsim\D\ve_{\mbf M_2'}-\D\ve_{\mbf M_1'}$.}
\end{figure}

Recent experiments on the kagome metals $\txt{AV}_3\txt{Sb}_5$ with $\txt{A}=\txt{K},\txt{Rb},\txt{Cs}$ have uncovered the relevance of topological charge density waves on the $\txt{V}$ kagome lattices \cite{ortiz19prm,yang20sa,ortiz20prl,kenney21jpcm,jiang20ax,yu21prb,zhao21ax,liang21ax,uykur21ax,chen21axpdw,li21ax}. These states develop at $80\txt{--}110\txt{ K}$ far above the superconductivity at $0.9\txt{--}2.7\txt{ K}$, with the charge modulations occurring at the three nesting momenta $\mbf Q_\a$. The observed giant anomalous Hall effect suggests that the charge density waves may break the time-reversal symmetry spontaneously and become topological \cite{yang20sa,yu21prb}. Given the proximity of Fermi surface to the Van Hove singularity and the commensurate momenta, we propose that the topological charge density waves are realized by the $3Q$ complex orders in our theoretical model. 

The Fermi surface of $\txt{AV}_3\txt{Sb}_5$ lies on the particle bands on the $\txt{V}$ kagome lattices. With the total phase condition demanded, the complex orders at the three momenta carry unequal phases generally. This suggests the general breakdown of rotation symmetry in the $3Q$ complex charge density waves in $\txt{AV}_3\txt{Sb}_5$. An experiment on $\txt{KV}_3\txt{Sb}_5$ uncovers the charge modulations with the ratios of strengths $\apx(3.1,0.9,3.8)$ at the three momenta, which is reversed under opposite external magnetic field \cite{jiang20ax}. Choosing the complex phases $(\pi[56/180],\pi[252/180],\pi[52/180])$ for the $3Q$ order, we obtain the magnitudes of real orders at the three momenta $(\cos^2\phi_1,\cos^2\phi_2,\cos^2\phi_3)=(0.313,0.095,0.379)$. This result is extremely close to the experimentally observed ratios. Furthermore, the real-space star-of-David pattern leads to a hexagon density modulation which matches the measured results in the experiment [Fig.~\ref{fig:exp}(a)]. Importantly, the state lies in the Chern insulator phase of the $3Q$ complex charge density waves. This explains naturally how the giant anomalous Hall effects occur in the experimental measurements \cite{yang20sa,yu21prb}. Note that the gap structure may be affected by the deviations from the ideal Van Hove fermiology. For example, residual Fermi surfaces may appear around the valleys $\mbf M'_\a$ under finite doping or nonperfect nesting. In the multiband structure of $\txt{AV}_3\txt{Sb}_5$, the normal bands immune to charge density waves may also contribute to the residual Fermi surfaces. These effects may explain the (nearly) gapless signals in the momentum-dependent gaps, as well as the nonquantized parts in the measured anomalous Hall effects.

In addition to the $3Q$ charge density waves at $\mbf M_\a$, recent experiments on $\txt{CsV}_3\txt{Sb}_5$ also observe a $1Q$ charge density wave at a half momentum $\mbf M_3'$ and a rotation symmetry breaking along the same direction \cite{zhao21ax,chen21axpdw}. Here we propose an explanation also based on the theoretical model of $3Q$ complex charge density waves. Adopting the experimental ratios of strengths in the charge modulations $\apx(2,3,1)$ \cite{zhao21ax}, we assign the complex phases $(\pi[60/180],\pi[231/180],\pi[69/180])$ to the $3Q$ order. Three valleys with anisotropic gaps $\D\ve_{\mbf M_3'}<\D\ve_{\mbf M_1'}<\D\ve_{\mbf M_2'}$ are observed near the Fermi level [Fig.~\ref{fig:exp}(b)], where residual Fermi surfaces may appear under finite doping or nonperfect nesting. The low-energy valleys may support secondary orders at low temperatures. If a secondary nematic order forms at the minimal-gap valley $\mbf M_3'$, rotation symmetry breaking occurs along this direction. An intervalley charge density wave may develop between the rest two valleys $\mbf M_{1,2}'$ and manifest a $1Q$ density modulation at the momentum $\mbf M_3'$. These valleys may also support the pair density waves at the three $3\mbf M_\a/4$, which were recently proposed based on the experiments \cite{chen21axpdw}. While we have chosen the state in the Chern insulator phase, the actual phases in the experiments can be examined by the measurement of anomalous Hall effect.

\section{Discussion}

We have studied the interplay of real and imaginary charge density waves at the Van Hove singularity on the hexagonal lattices. The $3Q$ complex orders constitute a rich Haldane-model phase diagram under a total phase condition, where rotation symmetry breaking occurs generally. When effective time-reversal symmetries arise under 1-site translations, the Dirac semimetals are protected. The breakdown of these symmetries gaps the Dirac points and leads to the trivial and Chern insulator phases. The exotic single-Dirac-point semimetals also appear along the gapless phase boundary. The theoretical model offers transparent interpretations to the experimentally observed topological charge density waves in the kagome metals $\txt{AV}_3\txt{Sb}_5$ with $\txt{A}=\txt{K},\txt{Rb},\txt{Cs}$, as well as the lower-temperature symmetry-breaking phenomena. Future experimental results may also find useful hints from the theoretical model in our analysis. Feasible experimental probes of complex charge density waves include the real-space microscopy and transport measurement. In addition to the solid-state materials, the ultracold atomic systems have also served as fertile grounds for the topological phases \cite{cooper19rmp}. The realization of tunable $3Q$ complex charge density waves may serve as an exciting direction for experimental investigations.

Our analysis has focused on the irreducible pairing channels under the symmetries, while strong deviations may occur away from the Fermi surface. We have addressed this issue briefly and obtained the qualitative results by considering the mixing between different real orders. Meanwhile, the finite doping or nonperfect nesting may also alter the gap structures. Further analysis in the practical systems can offer more precise predictions in the gap structures and real-space patterns, which is an interesting topic for future work. On the other hand, later analysis uncovers the possible occurrence of higher-order topological insulators in the $3Q$ charge bond orders \cite{lin21axhoti}. Inspecting such an unconventional state in the Haldane-model phase diagram is an interesting topic for future work.

Our work presents a complete framework of charge density waves at the $M$-point Van Hove singularity on the hexagonal lattices. While the model has presented transparent interpretations to the kagome metals $\txt{AV}_3\txt{Sb}_5$, the application to the other hexagonal lattice materials, such as the moir\'e systems \cite{lin19prb}, may also uncover intriguing phenomena. Note that the charge density waves may become incommensurate when the saddle points are shifted away from the $M$ points. The deformation of bands under such deviation may lead to nonperfect nesting and according residual Fermi surfaces. Meanwhile, the two decoupled channels both become complex \cite{lin19prb}, where the sliding phases may contribute to additional transport signals. The investigations along this direction may serve as an important topic for future work.

Our work also serves as a paradigmatic example of how the real and imaginary orders may interplay generally. The analysis herein may be generalized to the study of the other channels. The spin density waves at the Van Hove singularity on the hexagonal lattices may serve as an interesting example \cite{venderbos16prbsdw}. While the real orders realize the chiral noncoplanar Chern insulator \cite{martin08prl} and uniaxial half metal \cite{nandkishore12prl}, the imaginary order may support a quantum spin Hall insulator \cite{venderbos16prbsdw}. The interplay between these topological states can lead to an unconventional phase diagram of complex spin density waves. Such an analysis also presents an interesting topic for future work.

{\it Note added.} Recently, we learned about the independent mean-field studies on the charge density waves in the kagome metals $\txt{AV}_3\txt{Sb}_5$. A manuscript evaluated the mean-field ground state energies of various density, bond, and flux orders with mixed real and imaginary orders \cite{feng21ax}. Our analysis treats the real and imaginary orders as decoupled channels properly, thereby providing a more complete investigation of the interplay between these two orders. The actual energetically favored ground states are determined accordingly. Another manuscript conducted a Ginzburg-Landau analysis for another type of complex charge density wave \cite{denner21ax}. The real and imaginary orders therein differ from the $M$-point irreducible pairing channels discussed in our analysis. Meanwhile, our work presents a detailed analysis of the phase diagram and according ground states. A matching to the experimental results is further presented in our work.

\begin{acknowledgments}
This research was sponsored by the Army Research Office and was accomplished under Grant No. W911NF-17-1-0482. The views and conclusions contained in this document are those of the authors and should not be interpreted as representing the official policies, either expressed or implied, of the Army Research Office or the U.S. Government. The U.S. Government is authorized to reproduce and distribute reprints for Government purposes notwithstanding any copyright notation herein.
\end{acknowledgments}




\bibliography{Reference}

\begin{thebibliography}{58}%
\makeatletter
\providecommand \@ifxundefined [1]{%
 \@ifx{#1\undefined}
}%
\providecommand \@ifnum [1]{%
 \ifnum #1\expandafter \@firstoftwo
 \else \expandafter \@secondoftwo
 \fi
}%
\providecommand \@ifx [1]{%
 \ifx #1\expandafter \@firstoftwo
 \else \expandafter \@secondoftwo
 \fi
}%
\providecommand \natexlab [1]{#1}%
\providecommand \enquote  [1]{``#1''}%
\providecommand \bibnamefont  [1]{#1}%
\providecommand \bibfnamefont [1]{#1}%
\providecommand \citenamefont [1]{#1}%
\providecommand \href@noop [0]{\@secondoftwo}%
\providecommand \href [0]{\begingroup \@sanitize@url \@href}%
\providecommand \@href[1]{\@@startlink{#1}\@@href}%
\providecommand \@@href[1]{\endgroup#1\@@endlink}%
\providecommand \@sanitize@url [0]{\catcode `\\12\catcode `\$12\catcode
  `\&12\catcode `\#12\catcode `\^12\catcode `\_12\catcode `\%12\relax}%
\providecommand \@@startlink[1]{}%
\providecommand \@@endlink[0]{}%
\providecommand \url  [0]{\begingroup\@sanitize@url \@url }%
\providecommand \@url [1]{\endgroup\@href {#1}{\urlprefix }}%
\providecommand \urlprefix  [0]{URL }%
\providecommand \Eprint [0]{\href }%
\providecommand \doibase [0]{http://dx.doi.org/}%
\providecommand \selectlanguage [0]{\@gobble}%
\providecommand \bibinfo  [0]{\@secondoftwo}%
\providecommand \bibfield  [0]{\@secondoftwo}%
\providecommand \translation [1]{[#1]}%
\providecommand \BibitemOpen [0]{}%
\providecommand \bibitemStop [0]{}%
\providecommand \bibitemNoStop [0]{.\EOS\space}%
\providecommand \EOS [0]{\spacefactor3000\relax}%
\providecommand \BibitemShut  [1]{\csname bibitem#1\endcsname}%
\let\auto@bib@innerbib\@empty
\bibitem [{\citenamefont {Gonz\'alez}(2008)}]{gonzalez08prb}%
  \BibitemOpen
  \bibfield  {author} {\bibinfo {author} {\bibfnamefont {J.}~\bibnamefont
  {Gonz\'alez}},\ }\bibfield  {title} {\enquote {\bibinfo {title}
  {Kohn-luttinger superconductivity in graphene},}\ }\href {\doibase
  10.1103/PhysRevB.78.205431} {\bibfield  {journal} {\bibinfo  {journal} {Phys.
  Rev. B}\ }\textbf {\bibinfo {volume} {78}},\ \bibinfo {pages} {205431}
  (\bibinfo {year} {2008})}\BibitemShut {NoStop}%
\bibitem [{\citenamefont {Nandkishore}\ \emph
  {et~al.}(2012{\natexlab{a}})\citenamefont {Nandkishore}, \citenamefont
  {Levitov},\ and\ \citenamefont {Chubukov}}]{nandkishore12np}%
  \BibitemOpen
  \bibfield  {author} {\bibinfo {author} {\bibfnamefont {R.}~\bibnamefont
  {Nandkishore}}, \bibinfo {author} {\bibfnamefont {L.~S.}\ \bibnamefont
  {Levitov}}, \ and\ \bibinfo {author} {\bibfnamefont {A.~V.}\ \bibnamefont
  {Chubukov}},\ }\bibfield  {title} {\enquote {\bibinfo {title} {Chiral
  superconductivity from repulsive interactions in doped graphene},}\ }\href
  {\doibase 10.1038/nphys2208} {\bibfield  {journal} {\bibinfo  {journal} {Nat.
  Phys.}\ }\textbf {\bibinfo {volume} {8}},\ \bibinfo {pages} {158} (\bibinfo
  {year} {2012}{\natexlab{a}})}\BibitemShut {NoStop}%
\bibitem [{\citenamefont {Nandkishore}\ \emph
  {et~al.}(2012{\natexlab{b}})\citenamefont {Nandkishore}, \citenamefont
  {Chern},\ and\ \citenamefont {Chubukov}}]{nandkishore12prl}%
  \BibitemOpen
  \bibfield  {author} {\bibinfo {author} {\bibfnamefont {R.}~\bibnamefont
  {Nandkishore}}, \bibinfo {author} {\bibfnamefont {G.-W.}\ \bibnamefont
  {Chern}}, \ and\ \bibinfo {author} {\bibfnamefont {A.~V.}\ \bibnamefont
  {Chubukov}},\ }\bibfield  {title} {\enquote {\bibinfo {title} {Itinerant
  half-metal spin-density-wave state on the hexagonal lattice},}\ }\href
  {\doibase 10.1103/PhysRevLett.108.227204} {\bibfield  {journal} {\bibinfo
  {journal} {Phys. Rev. Lett.}\ }\textbf {\bibinfo {volume} {108}},\ \bibinfo
  {pages} {227204} (\bibinfo {year} {2012}{\natexlab{b}})}\BibitemShut
  {NoStop}%
\bibitem [{\citenamefont {Wang}\ \emph {et~al.}(2012)\citenamefont {Wang},
  \citenamefont {Xiang}, \citenamefont {Wang}, \citenamefont {Wang},
  \citenamefont {Yang},\ and\ \citenamefont {Lee}}]{wang12prb}%
  \BibitemOpen
  \bibfield  {author} {\bibinfo {author} {\bibfnamefont {W.-S.}\ \bibnamefont
  {Wang}}, \bibinfo {author} {\bibfnamefont {Y.-Y.}\ \bibnamefont {Xiang}},
  \bibinfo {author} {\bibfnamefont {Q.-H.}\ \bibnamefont {Wang}}, \bibinfo
  {author} {\bibfnamefont {F.}~\bibnamefont {Wang}}, \bibinfo {author}
  {\bibfnamefont {F.}~\bibnamefont {Yang}}, \ and\ \bibinfo {author}
  {\bibfnamefont {D.-H.}\ \bibnamefont {Lee}},\ }\bibfield  {title} {\enquote
  {\bibinfo {title} {Functional renormalization group and variational monte
  carlo studies of the electronic instabilities in graphene near $\frac{1}{4}$
  doping},}\ }\href {\doibase 10.1103/PhysRevB.85.035414} {\bibfield  {journal}
  {\bibinfo  {journal} {Phys. Rev. B}\ }\textbf {\bibinfo {volume} {85}},\
  \bibinfo {pages} {035414} (\bibinfo {year} {2012})}\BibitemShut {NoStop}%
\bibitem [{\citenamefont {Kiesel}\ \emph {et~al.}(2012)\citenamefont {Kiesel},
  \citenamefont {Platt}, \citenamefont {Hanke}, \citenamefont {Abanin},\ and\
  \citenamefont {Thomale}}]{kiesl12prb}%
  \BibitemOpen
  \bibfield  {author} {\bibinfo {author} {\bibfnamefont {M.~L.}\ \bibnamefont
  {Kiesel}}, \bibinfo {author} {\bibfnamefont {C.}~\bibnamefont {Platt}},
  \bibinfo {author} {\bibfnamefont {W.}~\bibnamefont {Hanke}}, \bibinfo
  {author} {\bibfnamefont {D.~A.}\ \bibnamefont {Abanin}}, \ and\ \bibinfo
  {author} {\bibfnamefont {R.}~\bibnamefont {Thomale}},\ }\bibfield  {title}
  {\enquote {\bibinfo {title} {Competing many-body instabilities and
  unconventional superconductivity in graphene},}\ }\href {\doibase
  10.1103/PhysRevB.86.020507} {\bibfield  {journal} {\bibinfo  {journal} {Phys.
  Rev. B}\ }\textbf {\bibinfo {volume} {86}},\ \bibinfo {pages} {020507}
  (\bibinfo {year} {2012})}\BibitemShut {NoStop}%
\bibitem [{\citenamefont {Jiang}\ \emph {et~al.}(2014)\citenamefont {Jiang},
  \citenamefont {Mesaros},\ and\ \citenamefont {Ran}}]{jiang14prx}%
  \BibitemOpen
  \bibfield  {author} {\bibinfo {author} {\bibfnamefont {S.}~\bibnamefont
  {Jiang}}, \bibinfo {author} {\bibfnamefont {A.}~\bibnamefont {Mesaros}}, \
  and\ \bibinfo {author} {\bibfnamefont {Y.}~\bibnamefont {Ran}},\ }\bibfield
  {title} {\enquote {\bibinfo {title} {Chiral spin-density wave,
  spin-charge-chern liquid, and $d+id$ superconductivity in $1/4$-doped
  correlated electronic systems on the honeycomb lattice},}\ }\href {\doibase
  10.1103/PhysRevX.4.031040} {\bibfield  {journal} {\bibinfo  {journal} {Phys.
  Rev. X}\ }\textbf {\bibinfo {volume} {4}},\ \bibinfo {pages} {031040}
  (\bibinfo {year} {2014})}\BibitemShut {NoStop}%
\bibitem [{\citenamefont {Nandkishore}\ \emph {et~al.}(2014)\citenamefont
  {Nandkishore}, \citenamefont {Thomale},\ and\ \citenamefont
  {Chubukov}}]{ntc}%
  \BibitemOpen
  \bibfield  {author} {\bibinfo {author} {\bibfnamefont {R.}~\bibnamefont
  {Nandkishore}}, \bibinfo {author} {\bibfnamefont {R.}~\bibnamefont
  {Thomale}}, \ and\ \bibinfo {author} {\bibfnamefont {A.~V.}\ \bibnamefont
  {Chubukov}},\ }\bibfield  {title} {\enquote {\bibinfo {title}
  {Superconductivity from weak repulsion in hexagonal lattice systems},}\
  }\href {\doibase 10.1103/PhysRevB.89.144501} {\bibfield  {journal} {\bibinfo
  {journal} {Phys. Rev. B}\ }\textbf {\bibinfo {volume} {89}},\ \bibinfo
  {pages} {144501} (\bibinfo {year} {2014})}\BibitemShut {NoStop}%
\bibitem [{\citenamefont {Martin}\ and\ \citenamefont
  {Batista}(2008)}]{martin08prl}%
  \BibitemOpen
  \bibfield  {author} {\bibinfo {author} {\bibfnamefont {I.}~\bibnamefont
  {Martin}}\ and\ \bibinfo {author} {\bibfnamefont {C.~D.}\ \bibnamefont
  {Batista}},\ }\bibfield  {title} {\enquote {\bibinfo {title} {Itinerant
  electron-driven chiral magnetic ordering and spontaneous quantum hall effect
  in triangular lattice models},}\ }\href {\doibase
  10.1103/PhysRevLett.101.156402} {\bibfield  {journal} {\bibinfo  {journal}
  {Phys. Rev. Lett.}\ }\textbf {\bibinfo {volume} {101}},\ \bibinfo {pages}
  {156402} (\bibinfo {year} {2008})}\BibitemShut {NoStop}%
\bibitem [{\citenamefont {Akagi}\ and\ \citenamefont
  {Motome}(2010)}]{akagi10jpsj}%
  \BibitemOpen
  \bibfield  {author} {\bibinfo {author} {\bibfnamefont {Y.}~\bibnamefont
  {Akagi}}\ and\ \bibinfo {author} {\bibfnamefont {Y.}~\bibnamefont {Motome}},\
  }\bibfield  {title} {\enquote {\bibinfo {title} {Spin chirality ordering and
  anomalous hall effect in the ferromagnetic kondo lattice model on a
  triangular lattice},}\ }\href {\doibase 10.1143/JPSJ.79.083711} {\bibfield
  {journal} {\bibinfo  {journal} {J. Phys. Soc. Jpn.}\ }\textbf {\bibinfo
  {volume} {79}},\ \bibinfo {pages} {083711} (\bibinfo {year}
  {2010})}\BibitemShut {NoStop}%
\bibitem [{\citenamefont {Akagi}\ \emph {et~al.}(2012)\citenamefont {Akagi},
  \citenamefont {Udagawa},\ and\ \citenamefont {Motome}}]{akagi12prl}%
  \BibitemOpen
  \bibfield  {author} {\bibinfo {author} {\bibfnamefont {Y.}~\bibnamefont
  {Akagi}}, \bibinfo {author} {\bibfnamefont {M.}~\bibnamefont {Udagawa}}, \
  and\ \bibinfo {author} {\bibfnamefont {Y.}~\bibnamefont {Motome}},\
  }\bibfield  {title} {\enquote {\bibinfo {title} {Hidden multiple-spin
  interactions as an origin of spin scalar chiral order in frustrated kondo
  lattice models},}\ }\href {\doibase 10.1103/PhysRevLett.108.096401}
  {\bibfield  {journal} {\bibinfo  {journal} {Phys. Rev. Lett.}\ }\textbf
  {\bibinfo {volume} {108}},\ \bibinfo {pages} {096401} (\bibinfo {year}
  {2012})}\BibitemShut {NoStop}%
\bibitem [{\citenamefont {Tieleman}\ \emph {et~al.}(2013)\citenamefont
  {Tieleman}, \citenamefont {Dutta}, \citenamefont {Lewenstein},\ and\
  \citenamefont {Eckardt}}]{tieleman13prl}%
  \BibitemOpen
  \bibfield  {author} {\bibinfo {author} {\bibfnamefont {O.}~\bibnamefont
  {Tieleman}}, \bibinfo {author} {\bibfnamefont {O.}~\bibnamefont {Dutta}},
  \bibinfo {author} {\bibfnamefont {M.}~\bibnamefont {Lewenstein}}, \ and\
  \bibinfo {author} {\bibfnamefont {A.}~\bibnamefont {Eckardt}},\ }\bibfield
  {title} {\enquote {\bibinfo {title} {Spontaneous time-reversal symmetry
  breaking for spinless fermions on a triangular lattice},}\ }\href {\doibase
  10.1103/PhysRevLett.110.096405} {\bibfield  {journal} {\bibinfo  {journal}
  {Phys. Rev. Lett.}\ }\textbf {\bibinfo {volume} {110}},\ \bibinfo {pages}
  {096405} (\bibinfo {year} {2013})}\BibitemShut {NoStop}%
\bibitem [{\citenamefont {Maharaj}\ \emph {et~al.}(2013)\citenamefont
  {Maharaj}, \citenamefont {Thomale},\ and\ \citenamefont
  {Raghu}}]{maharaj13prb}%
  \BibitemOpen
  \bibfield  {author} {\bibinfo {author} {\bibfnamefont {A.~V.}\ \bibnamefont
  {Maharaj}}, \bibinfo {author} {\bibfnamefont {R.}~\bibnamefont {Thomale}}, \
  and\ \bibinfo {author} {\bibfnamefont {S.}~\bibnamefont {Raghu}},\ }\bibfield
   {title} {\enquote {\bibinfo {title} {Particle-hole condensates of higher
  angular momentum in hexagonal systems},}\ }\href {\doibase
  10.1103/PhysRevB.88.205121} {\bibfield  {journal} {\bibinfo  {journal} {Phys.
  Rev. B}\ }\textbf {\bibinfo {volume} {88}},\ \bibinfo {pages} {205121}
  (\bibinfo {year} {2013})}\BibitemShut {NoStop}%
\bibitem [{\citenamefont {Yu}\ and\ \citenamefont {Li}(2012)}]{yu12prb}%
  \BibitemOpen
  \bibfield  {author} {\bibinfo {author} {\bibfnamefont {S.-L.}\ \bibnamefont
  {Yu}}\ and\ \bibinfo {author} {\bibfnamefont {J.-X.}\ \bibnamefont {Li}},\
  }\bibfield  {title} {\enquote {\bibinfo {title} {Chiral superconducting phase
  and chiral spin-density-wave phase in a hubbard model on the kagome
  lattice},}\ }\href {\doibase 10.1103/PhysRevB.85.144402} {\bibfield
  {journal} {\bibinfo  {journal} {Phys. Rev. B}\ }\textbf {\bibinfo {volume}
  {85}},\ \bibinfo {pages} {144402} (\bibinfo {year} {2012})}\BibitemShut
  {NoStop}%
\bibitem [{\citenamefont {Kiesel}\ \emph {et~al.}(2013)\citenamefont {Kiesel},
  \citenamefont {Platt},\ and\ \citenamefont {Thomale}}]{kiesel13prl}%
  \BibitemOpen
  \bibfield  {author} {\bibinfo {author} {\bibfnamefont {M.~L.}\ \bibnamefont
  {Kiesel}}, \bibinfo {author} {\bibfnamefont {C.}~\bibnamefont {Platt}}, \
  and\ \bibinfo {author} {\bibfnamefont {R.}~\bibnamefont {Thomale}},\
  }\bibfield  {title} {\enquote {\bibinfo {title} {Unconventional fermi surface
  instabilities in the kagome hubbard model},}\ }\href {\doibase
  10.1103/PhysRevLett.110.126405} {\bibfield  {journal} {\bibinfo  {journal}
  {Phys. Rev. Lett.}\ }\textbf {\bibinfo {volume} {110}},\ \bibinfo {pages}
  {126405} (\bibinfo {year} {2013})}\BibitemShut {NoStop}%
\bibitem [{\citenamefont {Wang}\ \emph {et~al.}(2013)\citenamefont {Wang},
  \citenamefont {Li}, \citenamefont {Xiang},\ and\ \citenamefont
  {Wang}}]{wang13prb}%
  \BibitemOpen
  \bibfield  {author} {\bibinfo {author} {\bibfnamefont {W.-S.}\ \bibnamefont
  {Wang}}, \bibinfo {author} {\bibfnamefont {Z.-Z.}\ \bibnamefont {Li}},
  \bibinfo {author} {\bibfnamefont {Y.-Y.}\ \bibnamefont {Xiang}}, \ and\
  \bibinfo {author} {\bibfnamefont {Q.-H.}\ \bibnamefont {Wang}},\ }\bibfield
  {title} {\enquote {\bibinfo {title} {Competing electronic orders on kagome
  lattices at van hove filling},}\ }\href {\doibase 10.1103/PhysRevB.87.115135}
  {\bibfield  {journal} {\bibinfo  {journal} {Phys. Rev. B}\ }\textbf {\bibinfo
  {volume} {87}},\ \bibinfo {pages} {115135} (\bibinfo {year}
  {2013})}\BibitemShut {NoStop}%
\bibitem [{\citenamefont {Van~Hove}(1953)}]{vanhove53pr}%
  \BibitemOpen
  \bibfield  {author} {\bibinfo {author} {\bibfnamefont {L.}~\bibnamefont
  {Van~Hove}},\ }\bibfield  {title} {\enquote {\bibinfo {title} {The occurrence
  of singularities in the elastic frequency distribution of a crystal},}\
  }\href {\doibase 10.1103/PhysRev.89.1189} {\bibfield  {journal} {\bibinfo
  {journal} {Phys. Rev.}\ }\textbf {\bibinfo {volume} {89}},\ \bibinfo {pages}
  {1189} (\bibinfo {year} {1953})}\BibitemShut {NoStop}%
\bibitem [{\citenamefont {McMillan}(1975)}]{mcmillan75prb}%
  \BibitemOpen
  \bibfield  {author} {\bibinfo {author} {\bibfnamefont {W.~L.}\ \bibnamefont
  {McMillan}},\ }\bibfield  {title} {\enquote {\bibinfo {title} {Landau theory
  of charge-density waves in transition-metal dichalcogenides},}\ }\href
  {\doibase 10.1103/PhysRevB.12.1187} {\bibfield  {journal} {\bibinfo
  {journal} {Phys. Rev. B}\ }\textbf {\bibinfo {volume} {12}},\ \bibinfo
  {pages} {1187} (\bibinfo {year} {1975})}\BibitemShut {NoStop}%
\bibitem [{\citenamefont {Ishioka}\ \emph {et~al.}(2010)\citenamefont
  {Ishioka}, \citenamefont {Liu}, \citenamefont {Shimatake}, \citenamefont
  {Kurosawa}, \citenamefont {Ichimura}, \citenamefont {Toda}, \citenamefont
  {Oda},\ and\ \citenamefont {Tanda}}]{ishioka10prl}%
  \BibitemOpen
  \bibfield  {author} {\bibinfo {author} {\bibfnamefont {J.}~\bibnamefont
  {Ishioka}}, \bibinfo {author} {\bibfnamefont {Y.~H.}\ \bibnamefont {Liu}},
  \bibinfo {author} {\bibfnamefont {K.}~\bibnamefont {Shimatake}}, \bibinfo
  {author} {\bibfnamefont {T.}~\bibnamefont {Kurosawa}}, \bibinfo {author}
  {\bibfnamefont {K.}~\bibnamefont {Ichimura}}, \bibinfo {author}
  {\bibfnamefont {Y.}~\bibnamefont {Toda}}, \bibinfo {author} {\bibfnamefont
  {M.}~\bibnamefont {Oda}}, \ and\ \bibinfo {author} {\bibfnamefont
  {S.}~\bibnamefont {Tanda}},\ }\bibfield  {title} {\enquote {\bibinfo {title}
  {Chiral charge-density waves},}\ }\href {\doibase
  10.1103/PhysRevLett.105.176401} {\bibfield  {journal} {\bibinfo  {journal}
  {Phys. Rev. Lett.}\ }\textbf {\bibinfo {volume} {105}},\ \bibinfo {pages}
  {176401} (\bibinfo {year} {2010})}\BibitemShut {NoStop}%
\bibitem [{\citenamefont {van Wezel}(2011)}]{vanwezel11epl}%
  \BibitemOpen
  \bibfield  {author} {\bibinfo {author} {\bibfnamefont {J.}~\bibnamefont {van
  Wezel}},\ }\bibfield  {title} {\enquote {\bibinfo {title} {Chirality and
  orbital order in charge density waves},}\ }\href {\doibase
  10.1209/0295-5075/96/67011} {\bibfield  {journal} {\bibinfo  {journal} {{EPL}
  (Europhys. Lett.)}\ }\textbf {\bibinfo {volume} {96}},\ \bibinfo {pages}
  {67011} (\bibinfo {year} {2011})}\BibitemShut {NoStop}%
\bibitem [{\citenamefont {McChesney}\ \emph {et~al.}(2010)\citenamefont
  {McChesney}, \citenamefont {Bostwick}, \citenamefont {Ohta}, \citenamefont
  {Seyller}, \citenamefont {Horn}, \citenamefont {Gonz\'alez},\ and\
  \citenamefont {Rotenberg}}]{mcchesney10prl}%
  \BibitemOpen
  \bibfield  {author} {\bibinfo {author} {\bibfnamefont {J.~L.}\ \bibnamefont
  {McChesney}}, \bibinfo {author} {\bibfnamefont {A.}~\bibnamefont {Bostwick}},
  \bibinfo {author} {\bibfnamefont {T.}~\bibnamefont {Ohta}}, \bibinfo {author}
  {\bibfnamefont {T.}~\bibnamefont {Seyller}}, \bibinfo {author} {\bibfnamefont
  {K.}~\bibnamefont {Horn}}, \bibinfo {author} {\bibfnamefont {J.}~\bibnamefont
  {Gonz\'alez}}, \ and\ \bibinfo {author} {\bibfnamefont {E.}~\bibnamefont
  {Rotenberg}},\ }\bibfield  {title} {\enquote {\bibinfo {title} {Extended van
  hove singularity and superconducting instability in doped graphene},}\ }\href
  {\doibase 10.1103/PhysRevLett.104.136803} {\bibfield  {journal} {\bibinfo
  {journal} {Phys. Rev. Lett.}\ }\textbf {\bibinfo {volume} {104}},\ \bibinfo
  {pages} {136803} (\bibinfo {year} {2010})}\BibitemShut {NoStop}%
\bibitem [{\citenamefont {Yuan}\ \emph {et~al.}(2019)\citenamefont {Yuan},
  \citenamefont {Isobe},\ and\ \citenamefont {Fu}}]{yuan19nc}%
  \BibitemOpen
  \bibfield  {author} {\bibinfo {author} {\bibfnamefont {N.~F.~Q.}\
  \bibnamefont {Yuan}}, \bibinfo {author} {\bibfnamefont {H.}~\bibnamefont
  {Isobe}}, \ and\ \bibinfo {author} {\bibfnamefont {L.}~\bibnamefont {Fu}},\
  }\bibfield  {title} {\enquote {\bibinfo {title} {Magic of high-order van hove
  singularity},}\ }\href {\doibase 10.1038/s41467-019-13670-9} {\bibfield
  {journal} {\bibinfo  {journal} {Nat. Commun.}\ }\textbf {\bibinfo {volume}
  {10}},\ \bibinfo {pages} {5769} (\bibinfo {year} {2019})}\BibitemShut
  {NoStop}%
\bibitem [{\citenamefont {Gonz\'alez}(2013)}]{gonzalez13prb}%
  \BibitemOpen
  \bibfield  {author} {\bibinfo {author} {\bibfnamefont {J.}~\bibnamefont
  {Gonz\'alez}},\ }\bibfield  {title} {\enquote {\bibinfo {title} {Magnetic and
  kohn-luttinger instabilities near a van hove singularity: Monolayer versus
  twisted bilayer graphene},}\ }\href {\doibase 10.1103/PhysRevB.88.125434}
  {\bibfield  {journal} {\bibinfo  {journal} {Phys. Rev. B}\ }\textbf {\bibinfo
  {volume} {88}},\ \bibinfo {pages} {125434} (\bibinfo {year}
  {2013})}\BibitemShut {NoStop}%
\bibitem [{\citenamefont {Classen}\ \emph {et~al.}(2020)\citenamefont
  {Classen}, \citenamefont {Chubukov}, \citenamefont {Honerkamp},\ and\
  \citenamefont {Scherer}}]{classen20prb}%
  \BibitemOpen
  \bibfield  {author} {\bibinfo {author} {\bibfnamefont {L.}~\bibnamefont
  {Classen}}, \bibinfo {author} {\bibfnamefont {A.~V.}\ \bibnamefont
  {Chubukov}}, \bibinfo {author} {\bibfnamefont {C.}~\bibnamefont {Honerkamp}},
  \ and\ \bibinfo {author} {\bibfnamefont {M.~M.}\ \bibnamefont {Scherer}},\
  }\bibfield  {title} {\enquote {\bibinfo {title} {Competing orders at
  higher-order van hove points},}\ }\href {\doibase
  10.1103/PhysRevB.102.125141} {\bibfield  {journal} {\bibinfo  {journal}
  {Phys. Rev. B}\ }\textbf {\bibinfo {volume} {102}},\ \bibinfo {pages}
  {125141} (\bibinfo {year} {2020})}\BibitemShut {NoStop}%
\bibitem [{\citenamefont {Lin}\ and\ \citenamefont
  {Nandkishore}(2020)}]{lin20prb}%
  \BibitemOpen
  \bibfield  {author} {\bibinfo {author} {\bibfnamefont {Y.-P.}\ \bibnamefont
  {Lin}}\ and\ \bibinfo {author} {\bibfnamefont {R.~M.}\ \bibnamefont
  {Nandkishore}},\ }\bibfield  {title} {\enquote {\bibinfo {title} {Parquet
  renormalization group analysis of weak-coupling instabilities with multiple
  high-order van hove points inside the brillouin zone},}\ }\href {\doibase
  10.1103/PhysRevB.102.245122} {\bibfield  {journal} {\bibinfo  {journal}
  {Phys. Rev. B}\ }\textbf {\bibinfo {volume} {102}},\ \bibinfo {pages}
  {245122} (\bibinfo {year} {2020})}\BibitemShut {NoStop}%
\bibitem [{\citenamefont {Affleck}\ and\ \citenamefont
  {Marston}(1988)}]{affleck88prb}%
  \BibitemOpen
  \bibfield  {author} {\bibinfo {author} {\bibfnamefont {I.}~\bibnamefont
  {Affleck}}\ and\ \bibinfo {author} {\bibfnamefont {J.~B.}\ \bibnamefont
  {Marston}},\ }\bibfield  {title} {\enquote {\bibinfo {title} {Large-n limit
  of the heisenberg-hubbard model: Implications for high-${T}_{c}$
  superconductors},}\ }\href {\doibase 10.1103/PhysRevB.37.3774} {\bibfield
  {journal} {\bibinfo  {journal} {Phys. Rev. B}\ }\textbf {\bibinfo {volume}
  {37}},\ \bibinfo {pages} {3774} (\bibinfo {year} {1988})}\BibitemShut
  {NoStop}%
\bibitem [{\citenamefont {Nayak}(2000)}]{nayak00prb}%
  \BibitemOpen
  \bibfield  {author} {\bibinfo {author} {\bibfnamefont {C.}~\bibnamefont
  {Nayak}},\ }\bibfield  {title} {\enquote {\bibinfo {title} {Density-wave
  states of nonzero angular momentum},}\ }\href {\doibase
  10.1103/PhysRevB.62.4880} {\bibfield  {journal} {\bibinfo  {journal} {Phys.
  Rev. B}\ }\textbf {\bibinfo {volume} {62}},\ \bibinfo {pages} {4880}
  (\bibinfo {year} {2000})}\BibitemShut {NoStop}%
\bibitem [{\citenamefont {Chakravarty}\ \emph {et~al.}(2001)\citenamefont
  {Chakravarty}, \citenamefont {Laughlin}, \citenamefont {Morr},\ and\
  \citenamefont {Nayak}}]{chakravarty01prb}%
  \BibitemOpen
  \bibfield  {author} {\bibinfo {author} {\bibfnamefont {S.}~\bibnamefont
  {Chakravarty}}, \bibinfo {author} {\bibfnamefont {R.~B.}\ \bibnamefont
  {Laughlin}}, \bibinfo {author} {\bibfnamefont {D.~K.}\ \bibnamefont {Morr}},
  \ and\ \bibinfo {author} {\bibfnamefont {C.}~\bibnamefont {Nayak}},\
  }\bibfield  {title} {\enquote {\bibinfo {title} {Hidden order in the
  cuprates},}\ }\href {\doibase 10.1103/PhysRevB.63.094503} {\bibfield
  {journal} {\bibinfo  {journal} {Phys. Rev. B}\ }\textbf {\bibinfo {volume}
  {63}},\ \bibinfo {pages} {094503} (\bibinfo {year} {2001})}\BibitemShut
  {NoStop}%
\bibitem [{\citenamefont {Venderbos}(2016{\natexlab{a}})}]{venderbos16prbcdw}%
  \BibitemOpen
  \bibfield  {author} {\bibinfo {author} {\bibfnamefont {J.~W.~F.}\
  \bibnamefont {Venderbos}},\ }\bibfield  {title} {\enquote {\bibinfo {title}
  {Symmetry analysis of translational symmetry broken density waves:
  Application to hexagonal lattices in two dimensions},}\ }\href {\doibase
  10.1103/PhysRevB.93.115107} {\bibfield  {journal} {\bibinfo  {journal} {Phys.
  Rev. B}\ }\textbf {\bibinfo {volume} {93}},\ \bibinfo {pages} {115107}
  (\bibinfo {year} {2016}{\natexlab{a}})}\BibitemShut {NoStop}%
\bibitem [{\citenamefont {Lin}\ and\ \citenamefont
  {Nandkishore}(2019)}]{lin19prb}%
  \BibitemOpen
  \bibfield  {author} {\bibinfo {author} {\bibfnamefont {Y.-P.}\ \bibnamefont
  {Lin}}\ and\ \bibinfo {author} {\bibfnamefont {R.~M.}\ \bibnamefont
  {Nandkishore}},\ }\bibfield  {title} {\enquote {\bibinfo {title} {Chiral
  twist on the high-${T}_{c}$ phase diagram in moir\'e heterostructures},}\
  }\href {\doibase 10.1103/PhysRevB.100.085136} {\bibfield  {journal} {\bibinfo
   {journal} {Phys. Rev. B}\ }\textbf {\bibinfo {volume} {100}},\ \bibinfo
  {pages} {085136} (\bibinfo {year} {2019})}\BibitemShut {NoStop}%
\bibitem [{\citenamefont {Venderbos}(2016{\natexlab{b}})}]{venderbos16prbsdw}%
  \BibitemOpen
  \bibfield  {author} {\bibinfo {author} {\bibfnamefont {J.~W.~F.}\
  \bibnamefont {Venderbos}},\ }\bibfield  {title} {\enquote {\bibinfo {title}
  {Multi-$q$ hexagonal spin density waves and dynamically generated spin-orbit
  coupling: Time-reversal invariant analog of the chiral spin density wave},}\
  }\href {\doibase 10.1103/PhysRevB.93.115108} {\bibfield  {journal} {\bibinfo
  {journal} {Phys. Rev. B}\ }\textbf {\bibinfo {volume} {93}},\ \bibinfo
  {pages} {115108} (\bibinfo {year} {2016}{\natexlab{b}})}\BibitemShut
  {NoStop}%
\bibitem [{\citenamefont {Classen}\ \emph {et~al.}(2019)\citenamefont
  {Classen}, \citenamefont {Honerkamp},\ and\ \citenamefont
  {Scherer}}]{classen19prb}%
  \BibitemOpen
  \bibfield  {author} {\bibinfo {author} {\bibfnamefont {L.}~\bibnamefont
  {Classen}}, \bibinfo {author} {\bibfnamefont {C.}~\bibnamefont {Honerkamp}},
  \ and\ \bibinfo {author} {\bibfnamefont {M.~M.}\ \bibnamefont {Scherer}},\
  }\bibfield  {title} {\enquote {\bibinfo {title} {Competing phases of
  interacting electrons on triangular lattices in moir\'e heterostructures},}\
  }\href {\doibase 10.1103/PhysRevB.99.195120} {\bibfield  {journal} {\bibinfo
  {journal} {Phys. Rev. B}\ }\textbf {\bibinfo {volume} {99}},\ \bibinfo
  {pages} {195120} (\bibinfo {year} {2019})}\BibitemShut {NoStop}%
\bibitem [{\citenamefont {Song}\ \emph {et~al.}(2021)\citenamefont {Song},
  \citenamefont {Vishwanath},\ and\ \citenamefont {Zhang}}]{song21prb}%
  \BibitemOpen
  \bibfield  {author} {\bibinfo {author} {\bibfnamefont {X.-Y.}\ \bibnamefont
  {Song}}, \bibinfo {author} {\bibfnamefont {A.}~\bibnamefont {Vishwanath}}, \
  and\ \bibinfo {author} {\bibfnamefont {Y.-H.}\ \bibnamefont {Zhang}},\
  }\bibfield  {title} {\enquote {\bibinfo {title} {Doping the chiral spin
  liquid: Topological superconductor or chiral metal},}\ }\href {\doibase
  10.1103/PhysRevB.103.165138} {\bibfield  {journal} {\bibinfo  {journal}
  {Phys. Rev. B}\ }\textbf {\bibinfo {volume} {103}},\ \bibinfo {pages}
  {165138} (\bibinfo {year} {2021})}\BibitemShut {NoStop}%
\bibitem [{\citenamefont {Ortiz}\ \emph {et~al.}(2019)\citenamefont {Ortiz},
  \citenamefont {Gomes}, \citenamefont {Morey}, \citenamefont {Winiarski},
  \citenamefont {Bordelon}, \citenamefont {Mangum}, \citenamefont {Oswald},
  \citenamefont {Rodriguez-Rivera}, \citenamefont {Neilson}, \citenamefont
  {Wilson}, \citenamefont {Ertekin}, \citenamefont {McQueen},\ and\
  \citenamefont {Toberer}}]{ortiz19prm}%
  \BibitemOpen
  \bibfield  {author} {\bibinfo {author} {\bibfnamefont {B.~R.}\ \bibnamefont
  {Ortiz}}, \bibinfo {author} {\bibfnamefont {L.~C.}\ \bibnamefont {Gomes}},
  \bibinfo {author} {\bibfnamefont {J.~R.}\ \bibnamefont {Morey}}, \bibinfo
  {author} {\bibfnamefont {M.}~\bibnamefont {Winiarski}}, \bibinfo {author}
  {\bibfnamefont {M.}~\bibnamefont {Bordelon}}, \bibinfo {author}
  {\bibfnamefont {J.~S.}\ \bibnamefont {Mangum}}, \bibinfo {author}
  {\bibfnamefont {I.~W.~H.}\ \bibnamefont {Oswald}}, \bibinfo {author}
  {\bibfnamefont {J.~A.}\ \bibnamefont {Rodriguez-Rivera}}, \bibinfo {author}
  {\bibfnamefont {J.~R.}\ \bibnamefont {Neilson}}, \bibinfo {author}
  {\bibfnamefont {S.~D.}\ \bibnamefont {Wilson}}, \bibinfo {author}
  {\bibfnamefont {E.}~\bibnamefont {Ertekin}}, \bibinfo {author} {\bibfnamefont
  {T.~M.}\ \bibnamefont {McQueen}}, \ and\ \bibinfo {author} {\bibfnamefont
  {E.~S.}\ \bibnamefont {Toberer}},\ }\bibfield  {title} {\enquote {\bibinfo
  {title} {New kagome prototype materials: discovery of
  ${\mathrm{kv}}_{3}{\mathrm{sb}}_{5},{\mathrm{rbv}}_{3}{\mathrm{sb}}_{5}$, and
  ${\mathrm{csv}}_{3}{\mathrm{sb}}_{5}$},}\ }\href {\doibase
  10.1103/PhysRevMaterials.3.094407} {\bibfield  {journal} {\bibinfo  {journal}
  {Phys. Rev. Materials}\ }\textbf {\bibinfo {volume} {3}},\ \bibinfo {pages}
  {094407} (\bibinfo {year} {2019})}\BibitemShut {NoStop}%
\bibitem [{\citenamefont {Yang}\ \emph {et~al.}(2020)\citenamefont {Yang},
  \citenamefont {Wang}, \citenamefont {Ortiz}, \citenamefont {Liu},
  \citenamefont {Gayles}, \citenamefont {Derunova}, \citenamefont
  {Gonzalez-Hernandez}, \citenamefont {{\v S}mejkal}, \citenamefont {Chen},
  \citenamefont {Parkin}, \citenamefont {Wilson}, \citenamefont {Toberer},
  \citenamefont {McQueen},\ and\ \citenamefont {Ali}}]{yang20sa}%
  \BibitemOpen
  \bibfield  {author} {\bibinfo {author} {\bibfnamefont {S.-Y.}\ \bibnamefont
  {Yang}}, \bibinfo {author} {\bibfnamefont {Y.}~\bibnamefont {Wang}}, \bibinfo
  {author} {\bibfnamefont {B.~R.}\ \bibnamefont {Ortiz}}, \bibinfo {author}
  {\bibfnamefont {D.}~\bibnamefont {Liu}}, \bibinfo {author} {\bibfnamefont
  {J.}~\bibnamefont {Gayles}}, \bibinfo {author} {\bibfnamefont
  {E.}~\bibnamefont {Derunova}}, \bibinfo {author} {\bibfnamefont
  {R.}~\bibnamefont {Gonzalez-Hernandez}}, \bibinfo {author} {\bibfnamefont
  {L.}~\bibnamefont {{\v S}mejkal}}, \bibinfo {author} {\bibfnamefont
  {Y.}~\bibnamefont {Chen}}, \bibinfo {author} {\bibfnamefont {S.~S.~P.}\
  \bibnamefont {Parkin}}, \bibinfo {author} {\bibfnamefont {S.~D.}\
  \bibnamefont {Wilson}}, \bibinfo {author} {\bibfnamefont {E.~S.}\
  \bibnamefont {Toberer}}, \bibinfo {author} {\bibfnamefont {T.}~\bibnamefont
  {McQueen}}, \ and\ \bibinfo {author} {\bibfnamefont {M.~N.}\ \bibnamefont
  {Ali}},\ }\bibfield  {title} {\enquote {\bibinfo {title} {Giant,
  unconventional anomalous hall effect in the metallic frustrated magnet
  candidate, kv3sb5},}\ }\href {\doibase 10.1126/sciadv.abb6003} {\bibfield
  {journal} {\bibinfo  {journal} {Sci. Adv.}\ }\textbf {\bibinfo {volume}
  {6}},\ \bibinfo {pages} {eabb6003} (\bibinfo {year} {2020})}\BibitemShut
  {NoStop}%
\bibitem [{\citenamefont {Ortiz}\ \emph {et~al.}(2020)\citenamefont {Ortiz},
  \citenamefont {Teicher}, \citenamefont {Hu}, \citenamefont {Zuo},
  \citenamefont {Sarte}, \citenamefont {Schueller}, \citenamefont {Abeykoon},
  \citenamefont {Krogstad}, \citenamefont {Rosenkranz}, \citenamefont {Osborn},
  \citenamefont {Seshadri}, \citenamefont {Balents}, \citenamefont {He},\ and\
  \citenamefont {Wilson}}]{ortiz20prl}%
  \BibitemOpen
  \bibfield  {author} {\bibinfo {author} {\bibfnamefont {B.~R.}\ \bibnamefont
  {Ortiz}}, \bibinfo {author} {\bibfnamefont {S.~M.~L.}\ \bibnamefont
  {Teicher}}, \bibinfo {author} {\bibfnamefont {Y.}~\bibnamefont {Hu}},
  \bibinfo {author} {\bibfnamefont {J.~L.}\ \bibnamefont {Zuo}}, \bibinfo
  {author} {\bibfnamefont {P.~M.}\ \bibnamefont {Sarte}}, \bibinfo {author}
  {\bibfnamefont {E.~C.}\ \bibnamefont {Schueller}}, \bibinfo {author}
  {\bibfnamefont {A.~M.~M.}\ \bibnamefont {Abeykoon}}, \bibinfo {author}
  {\bibfnamefont {M.~J.}\ \bibnamefont {Krogstad}}, \bibinfo {author}
  {\bibfnamefont {S.}~\bibnamefont {Rosenkranz}}, \bibinfo {author}
  {\bibfnamefont {R.}~\bibnamefont {Osborn}}, \bibinfo {author} {\bibfnamefont
  {R.}~\bibnamefont {Seshadri}}, \bibinfo {author} {\bibfnamefont
  {L.}~\bibnamefont {Balents}}, \bibinfo {author} {\bibfnamefont
  {J.}~\bibnamefont {He}}, \ and\ \bibinfo {author} {\bibfnamefont {S.~D.}\
  \bibnamefont {Wilson}},\ }\bibfield  {title} {\enquote {\bibinfo {title}
  {$\mathrm{Cs}{\mathrm{v}}_{3}{\mathrm{sb}}_{5}$: A ${\mathbb{z}}_{2}$
  topological kagome metal with a superconducting ground state},}\ }\href
  {\doibase 10.1103/PhysRevLett.125.247002} {\bibfield  {journal} {\bibinfo
  {journal} {Phys. Rev. Lett.}\ }\textbf {\bibinfo {volume} {125}},\ \bibinfo
  {pages} {247002} (\bibinfo {year} {2020})}\BibitemShut {NoStop}%
\bibitem [{\citenamefont {Kenney}\ \emph {et~al.}(2021)\citenamefont {Kenney},
  \citenamefont {Ortiz}, \citenamefont {Wang}, \citenamefont {Wilson},\ and\
  \citenamefont {Graf}}]{kenney21jpcm}%
  \BibitemOpen
  \bibfield  {author} {\bibinfo {author} {\bibfnamefont {E.~M.}\ \bibnamefont
  {Kenney}}, \bibinfo {author} {\bibfnamefont {B.~R.}\ \bibnamefont {Ortiz}},
  \bibinfo {author} {\bibfnamefont {C.}~\bibnamefont {Wang}}, \bibinfo {author}
  {\bibfnamefont {S.~D.}\ \bibnamefont {Wilson}}, \ and\ \bibinfo {author}
  {\bibfnamefont {M.~J.}\ \bibnamefont {Graf}},\ }\bibfield  {title} {\enquote
  {\bibinfo {title} {Absence of local moments in the kagome metal {KV}3sb5 as
  determined by muon spin spectroscopy},}\ }\href {\doibase
  10.1088/1361-648x/abe8f9} {\bibfield  {journal} {\bibinfo  {journal} {J.
  Phys.: Condens. Matter}\ }\textbf {\bibinfo {volume} {33}},\ \bibinfo {pages}
  {235801} (\bibinfo {year} {2021})}\BibitemShut {NoStop}%
\bibitem [{\citenamefont {{Jiang}}\ \emph {et~al.}(2020)\citenamefont
  {{Jiang}}, \citenamefont {{Yin}}, \citenamefont {{Denner}}, \citenamefont
  {{Shumiya}}, \citenamefont {{Ortiz}}, \citenamefont {{Xu}}, \citenamefont
  {{Guguchia}}, \citenamefont {{He}}, \citenamefont {{Shafayat Hossain}},
  \citenamefont {{Liu}}, \citenamefont {{Ruff}}, \citenamefont {{Kautzsch}},
  \citenamefont {{Zhang}}, \citenamefont {{Chang}}, \citenamefont
  {{Belopolski}}, \citenamefont {{Zhang}}, \citenamefont {{Cochran}},
  \citenamefont {{Multer}}, \citenamefont {{Litskevich}}, \citenamefont
  {{Cheng}}, \citenamefont {{Yang}}, \citenamefont {{Wang}}, \citenamefont
  {{Thomale}}, \citenamefont {{Neupert}}, \citenamefont {{Wilson}},\ and\
  \citenamefont {{Zahid Hasan}}}]{jiang20ax}%
  \BibitemOpen
  \bibfield  {author} {\bibinfo {author} {\bibfnamefont {Y.-X.}\ \bibnamefont
  {{Jiang}}}, \bibinfo {author} {\bibfnamefont {J.-X.}\ \bibnamefont {{Yin}}},
  \bibinfo {author} {\bibfnamefont {M.~M.}\ \bibnamefont {{Denner}}}, \bibinfo
  {author} {\bibfnamefont {N.}~\bibnamefont {{Shumiya}}}, \bibinfo {author}
  {\bibfnamefont {B.~R.}\ \bibnamefont {{Ortiz}}}, \bibinfo {author}
  {\bibfnamefont {G.}~\bibnamefont {{Xu}}}, \bibinfo {author} {\bibfnamefont
  {Z.}~\bibnamefont {{Guguchia}}}, \bibinfo {author} {\bibfnamefont
  {J.}~\bibnamefont {{He}}}, \bibinfo {author} {\bibfnamefont {M.}~\bibnamefont
  {{Shafayat Hossain}}}, \bibinfo {author} {\bibfnamefont {X.}~\bibnamefont
  {{Liu}}}, \bibinfo {author} {\bibfnamefont {J.}~\bibnamefont {{Ruff}}},
  \bibinfo {author} {\bibfnamefont {L.}~\bibnamefont {{Kautzsch}}}, \bibinfo
  {author} {\bibfnamefont {S.~S.}\ \bibnamefont {{Zhang}}}, \bibinfo {author}
  {\bibfnamefont {G.}~\bibnamefont {{Chang}}}, \bibinfo {author} {\bibfnamefont
  {I.}~\bibnamefont {{Belopolski}}}, \bibinfo {author} {\bibfnamefont
  {Q.}~\bibnamefont {{Zhang}}}, \bibinfo {author} {\bibfnamefont {T.~A.}\
  \bibnamefont {{Cochran}}}, \bibinfo {author} {\bibfnamefont {D.}~\bibnamefont
  {{Multer}}}, \bibinfo {author} {\bibfnamefont {M.}~\bibnamefont
  {{Litskevich}}}, \bibinfo {author} {\bibfnamefont {Z.-J.}\ \bibnamefont
  {{Cheng}}}, \bibinfo {author} {\bibfnamefont {X.~P.}\ \bibnamefont {{Yang}}},
  \bibinfo {author} {\bibfnamefont {Z.}~\bibnamefont {{Wang}}}, \bibinfo
  {author} {\bibfnamefont {R.}~\bibnamefont {{Thomale}}}, \bibinfo {author}
  {\bibfnamefont {T.}~\bibnamefont {{Neupert}}}, \bibinfo {author}
  {\bibfnamefont {S.~D.}\ \bibnamefont {{Wilson}}}, \ and\ \bibinfo {author}
  {\bibfnamefont {M.}~\bibnamefont {{Zahid Hasan}}},\ }\bibfield  {title}
  {\enquote {\bibinfo {title} {{Discovery of unconventional chiral charge order
  in kagome superconductor KV3Sb5}},}\ }\href@noop {} {\bibfield  {journal}
  {\bibinfo  {journal} {arXiv e-prints}\ ,\ \bibinfo {eid} {arXiv:2012.15709}}
  (\bibinfo {year} {2020})},\ \Eprint {http://arxiv.org/abs/2012.15709}
  {arXiv:2012.15709 [cond-mat.supr-con]} \BibitemShut {NoStop}%
\bibitem [{\citenamefont {Yu}\ \emph {et~al.}(2021)\citenamefont {Yu},
  \citenamefont {Wu}, \citenamefont {Wang}, \citenamefont {Lei}, \citenamefont
  {Zhuo}, \citenamefont {Ying},\ and\ \citenamefont {Chen}}]{yu21prb}%
  \BibitemOpen
  \bibfield  {author} {\bibinfo {author} {\bibfnamefont {F.~H.}\ \bibnamefont
  {Yu}}, \bibinfo {author} {\bibfnamefont {T.}~\bibnamefont {Wu}}, \bibinfo
  {author} {\bibfnamefont {Z.~Y.}\ \bibnamefont {Wang}}, \bibinfo {author}
  {\bibfnamefont {B.}~\bibnamefont {Lei}}, \bibinfo {author} {\bibfnamefont
  {W.~Z.}\ \bibnamefont {Zhuo}}, \bibinfo {author} {\bibfnamefont {J.~J.}\
  \bibnamefont {Ying}}, \ and\ \bibinfo {author} {\bibfnamefont {X.~H.}\
  \bibnamefont {Chen}},\ }\bibfield  {title} {\enquote {\bibinfo {title}
  {Concurrence of anomalous hall effect and charge density wave in a
  superconducting topological kagome metal},}\ }\href {\doibase
  10.1103/PhysRevB.104.L041103} {\bibfield  {journal} {\bibinfo  {journal}
  {Phys. Rev. B}\ }\textbf {\bibinfo {volume} {104}},\ \bibinfo {pages}
  {L041103} (\bibinfo {year} {2021})}\BibitemShut {NoStop}%
\bibitem [{\citenamefont {{Zhao}}\ \emph
  {et~al.}(2021{\natexlab{a}})\citenamefont {{Zhao}}, \citenamefont {{Li}},
  \citenamefont {{Ortiz}}, \citenamefont {{Teicher}}, \citenamefont {{Park}},
  \citenamefont {{Ye}}, \citenamefont {{Wang}}, \citenamefont {{Balents}},
  \citenamefont {{Wilson}},\ and\ \citenamefont {{Zeljkovic}}}]{zhao21ax}%
  \BibitemOpen
  \bibfield  {author} {\bibinfo {author} {\bibfnamefont {H.}~\bibnamefont
  {{Zhao}}}, \bibinfo {author} {\bibfnamefont {H.}~\bibnamefont {{Li}}},
  \bibinfo {author} {\bibfnamefont {B.~R.}\ \bibnamefont {{Ortiz}}}, \bibinfo
  {author} {\bibfnamefont {S.~M.~L.}\ \bibnamefont {{Teicher}}}, \bibinfo
  {author} {\bibfnamefont {T.}~\bibnamefont {{Park}}}, \bibinfo {author}
  {\bibfnamefont {M.}~\bibnamefont {{Ye}}}, \bibinfo {author} {\bibfnamefont
  {Z.}~\bibnamefont {{Wang}}}, \bibinfo {author} {\bibfnamefont
  {L.}~\bibnamefont {{Balents}}}, \bibinfo {author} {\bibfnamefont {S.~D.}\
  \bibnamefont {{Wilson}}}, \ and\ \bibinfo {author} {\bibfnamefont
  {I.}~\bibnamefont {{Zeljkovic}}},\ }\bibfield  {title} {\enquote {\bibinfo
  {title} {{Cascade of correlated electron states in a kagome superconductor
  CsV3Sb5}},}\ }\href@noop {} {\bibfield  {journal} {\bibinfo  {journal} {arXiv
  e-prints}\ ,\ \bibinfo {eid} {arXiv:2103.03118}} (\bibinfo {year}
  {2021}{\natexlab{a}})},\ \Eprint {http://arxiv.org/abs/2103.03118}
  {arXiv:2103.03118 [cond-mat.supr-con]} \BibitemShut {NoStop}%
\bibitem [{\citenamefont {{Liang}}\ \emph {et~al.}(2021)\citenamefont
  {{Liang}}, \citenamefont {{Hou}}, \citenamefont {{Ma}}, \citenamefont
  {{Zhang}}, \citenamefont {{Wu}}, \citenamefont {{Zhang}}, \citenamefont
  {{Yu}}, \citenamefont {{Ying}}, \citenamefont {{Jiang}}, \citenamefont
  {{Shan}}, \citenamefont {{Wang}},\ and\ \citenamefont {{Chen}}}]{liang21ax}%
  \BibitemOpen
  \bibfield  {author} {\bibinfo {author} {\bibfnamefont {Z.}~\bibnamefont
  {{Liang}}}, \bibinfo {author} {\bibfnamefont {X.}~\bibnamefont {{Hou}}},
  \bibinfo {author} {\bibfnamefont {W.}~\bibnamefont {{Ma}}}, \bibinfo {author}
  {\bibfnamefont {F.}~\bibnamefont {{Zhang}}}, \bibinfo {author} {\bibfnamefont
  {P.}~\bibnamefont {{Wu}}}, \bibinfo {author} {\bibfnamefont {Z.}~\bibnamefont
  {{Zhang}}}, \bibinfo {author} {\bibfnamefont {F.}~\bibnamefont {{Yu}}},
  \bibinfo {author} {\bibfnamefont {J.~J.}\ \bibnamefont {{Ying}}}, \bibinfo
  {author} {\bibfnamefont {K.}~\bibnamefont {{Jiang}}}, \bibinfo {author}
  {\bibfnamefont {L.}~\bibnamefont {{Shan}}}, \bibinfo {author} {\bibfnamefont
  {Z.}~\bibnamefont {{Wang}}}, \ and\ \bibinfo {author} {\bibfnamefont {X.~H.}\
  \bibnamefont {{Chen}}},\ }\bibfield  {title} {\enquote {\bibinfo {title}
  {{Three-dimensional charge density wave and robust zero-bias conductance peak
  inside the superconducting vortex core of a kagome superconductor
  CsV$_3$Sb$_5$}},}\ }\href@noop {} {\bibfield  {journal} {\bibinfo  {journal}
  {arXiv e-prints}\ ,\ \bibinfo {eid} {arXiv:2103.04760}} (\bibinfo {year}
  {2021})},\ \Eprint {http://arxiv.org/abs/2103.04760} {arXiv:2103.04760
  [cond-mat.supr-con]} \BibitemShut {NoStop}%
\bibitem [{\citenamefont {{Uykur}}\ \emph {et~al.}(2021)\citenamefont
  {{Uykur}}, \citenamefont {{Ortiz}}, \citenamefont {{Wilson}}, \citenamefont
  {{Dressel}},\ and\ \citenamefont {{Tsirlin}}}]{uykur21ax}%
  \BibitemOpen
  \bibfield  {author} {\bibinfo {author} {\bibfnamefont {E.}~\bibnamefont
  {{Uykur}}}, \bibinfo {author} {\bibfnamefont {B.~R.}\ \bibnamefont
  {{Ortiz}}}, \bibinfo {author} {\bibfnamefont {S.~D.}\ \bibnamefont
  {{Wilson}}}, \bibinfo {author} {\bibfnamefont {M.}~\bibnamefont {{Dressel}}},
  \ and\ \bibinfo {author} {\bibfnamefont {A.~A.}\ \bibnamefont {{Tsirlin}}},\
  }\bibfield  {title} {\enquote {\bibinfo {title} {{Optical detection of
  charge-density-wave instability in the non-magnetic kagome metal
  KV$_3$Sb$_5$}},}\ }\href@noop {} {\bibfield  {journal} {\bibinfo  {journal}
  {arXiv e-prints}\ ,\ \bibinfo {eid} {arXiv:2103.07912}} (\bibinfo {year}
  {2021})},\ \Eprint {http://arxiv.org/abs/2103.07912} {arXiv:2103.07912
  [cond-mat.str-el]} \BibitemShut {NoStop}%
\bibitem [{\citenamefont {{Chen}}\ \emph {et~al.}(2021)\citenamefont {{Chen}},
  \citenamefont {{Yang}}, \citenamefont {{Hu}}, \citenamefont {{Zhao}},
  \citenamefont {{Yuan}}, \citenamefont {{Xing}}, \citenamefont {{Qian}},
  \citenamefont {{Huang}}, \citenamefont {{Li}}, \citenamefont {{Ye}},
  \citenamefont {{Yin}}, \citenamefont {{Gong}}, \citenamefont {{Tu}},
  \citenamefont {{Lei}}, \citenamefont {{Ma}}, \citenamefont {{Zhang}},
  \citenamefont {{Ni}}, \citenamefont {{Tan}}, \citenamefont {{Shen}},
  \citenamefont {{Dong}}, \citenamefont {{Yan}}, \citenamefont {{Wang}},\ and\
  \citenamefont {{Gao}}}]{chen21axpdw}%
  \BibitemOpen
  \bibfield  {author} {\bibinfo {author} {\bibfnamefont {H.}~\bibnamefont
  {{Chen}}}, \bibinfo {author} {\bibfnamefont {H.}~\bibnamefont {{Yang}}},
  \bibinfo {author} {\bibfnamefont {B.}~\bibnamefont {{Hu}}}, \bibinfo {author}
  {\bibfnamefont {Z.}~\bibnamefont {{Zhao}}}, \bibinfo {author} {\bibfnamefont
  {J.}~\bibnamefont {{Yuan}}}, \bibinfo {author} {\bibfnamefont
  {Y.}~\bibnamefont {{Xing}}}, \bibinfo {author} {\bibfnamefont
  {G.}~\bibnamefont {{Qian}}}, \bibinfo {author} {\bibfnamefont
  {Z.}~\bibnamefont {{Huang}}}, \bibinfo {author} {\bibfnamefont
  {G.}~\bibnamefont {{Li}}}, \bibinfo {author} {\bibfnamefont {Y.}~\bibnamefont
  {{Ye}}}, \bibinfo {author} {\bibfnamefont {Q.}~\bibnamefont {{Yin}}},
  \bibinfo {author} {\bibfnamefont {C.}~\bibnamefont {{Gong}}}, \bibinfo
  {author} {\bibfnamefont {Z.}~\bibnamefont {{Tu}}}, \bibinfo {author}
  {\bibfnamefont {H.}~\bibnamefont {{Lei}}}, \bibinfo {author} {\bibfnamefont
  {S.}~\bibnamefont {{Ma}}}, \bibinfo {author} {\bibfnamefont {H.}~\bibnamefont
  {{Zhang}}}, \bibinfo {author} {\bibfnamefont {S.}~\bibnamefont {{Ni}}},
  \bibinfo {author} {\bibfnamefont {H.}~\bibnamefont {{Tan}}}, \bibinfo
  {author} {\bibfnamefont {C.}~\bibnamefont {{Shen}}}, \bibinfo {author}
  {\bibfnamefont {X.}~\bibnamefont {{Dong}}}, \bibinfo {author} {\bibfnamefont
  {B.}~\bibnamefont {{Yan}}}, \bibinfo {author} {\bibfnamefont
  {Z.}~\bibnamefont {{Wang}}}, \ and\ \bibinfo {author} {\bibfnamefont {H.-J.}\
  \bibnamefont {{Gao}}},\ }\bibfield  {title} {\enquote {\bibinfo {title}
  {{Roton pair density wave and unconventional strong-coupling
  superconductivity in a topological kagome metal}},}\ }\href@noop {}
  {\bibfield  {journal} {\bibinfo  {journal} {arXiv e-prints}\ ,\ \bibinfo
  {eid} {arXiv:2103.09188}} (\bibinfo {year} {2021})},\ \Eprint
  {http://arxiv.org/abs/2103.09188} {arXiv:2103.09188 [cond-mat.supr-con]}
  \BibitemShut {NoStop}%
\bibitem [{\citenamefont {{Li}}\ \emph {et~al.}(2021)\citenamefont {{Li}},
  \citenamefont {{Zhang}}, \citenamefont {{Pai}}, \citenamefont {{Marvinney}},
  \citenamefont {{Said}}, \citenamefont {{Yilmaz}}, \citenamefont {{Yin}},
  \citenamefont {{Gong}}, \citenamefont {{Tu}}, \citenamefont {{Vescovo}},
  \citenamefont {{Moore}}, \citenamefont {{Murakami}}, \citenamefont {{Lei}},
  \citenamefont {{Lee}}, \citenamefont {{Lawrie}},\ and\ \citenamefont
  {{Miao}}}]{li21ax}%
  \BibitemOpen
  \bibfield  {author} {\bibinfo {author} {\bibfnamefont {H.~X.}\ \bibnamefont
  {{Li}}}, \bibinfo {author} {\bibfnamefont {T.~T.}\ \bibnamefont {{Zhang}}},
  \bibinfo {author} {\bibfnamefont {Y.~Y.}\ \bibnamefont {{Pai}}}, \bibinfo
  {author} {\bibfnamefont {C.}~\bibnamefont {{Marvinney}}}, \bibinfo {author}
  {\bibfnamefont {A.}~\bibnamefont {{Said}}}, \bibinfo {author} {\bibfnamefont
  {T.}~\bibnamefont {{Yilmaz}}}, \bibinfo {author} {\bibfnamefont
  {Q.}~\bibnamefont {{Yin}}}, \bibinfo {author} {\bibfnamefont
  {C.}~\bibnamefont {{Gong}}}, \bibinfo {author} {\bibfnamefont
  {Z.}~\bibnamefont {{Tu}}}, \bibinfo {author} {\bibfnamefont {E.}~\bibnamefont
  {{Vescovo}}}, \bibinfo {author} {\bibfnamefont {R.~G.}\ \bibnamefont
  {{Moore}}}, \bibinfo {author} {\bibfnamefont {S.}~\bibnamefont {{Murakami}}},
  \bibinfo {author} {\bibfnamefont {H.~C.}\ \bibnamefont {{Lei}}}, \bibinfo
  {author} {\bibfnamefont {H.~N.}\ \bibnamefont {{Lee}}}, \bibinfo {author}
  {\bibfnamefont {B.}~\bibnamefont {{Lawrie}}}, \ and\ \bibinfo {author}
  {\bibfnamefont {H.}~\bibnamefont {{Miao}}},\ }\bibfield  {title} {\enquote
  {\bibinfo {title} {{Observation of Unconventional Charge Density Wave without
  Acoustic Phonon Anomaly in Kagome Superconductors AV3Sb5 (A=Rb,Cs)}},}\
  }\href@noop {} {\bibfield  {journal} {\bibinfo  {journal} {arXiv e-prints}\
  ,\ \bibinfo {eid} {arXiv:2103.09769}} (\bibinfo {year} {2021})},\ \Eprint
  {http://arxiv.org/abs/2103.09769} {arXiv:2103.09769 [cond-mat.supr-con]}
  \BibitemShut {NoStop}%
\bibitem [{\citenamefont {Ortiz}\ \emph {et~al.}(2021)\citenamefont {Ortiz},
  \citenamefont {Sarte}, \citenamefont {Kenney}, \citenamefont {Graf},
  \citenamefont {Teicher}, \citenamefont {Seshadri},\ and\ \citenamefont
  {Wilson}}]{ortiz21prm}%
  \BibitemOpen
  \bibfield  {author} {\bibinfo {author} {\bibfnamefont {B.~R.}\ \bibnamefont
  {Ortiz}}, \bibinfo {author} {\bibfnamefont {P.~M.}\ \bibnamefont {Sarte}},
  \bibinfo {author} {\bibfnamefont {E.~M.}\ \bibnamefont {Kenney}}, \bibinfo
  {author} {\bibfnamefont {M.~J.}\ \bibnamefont {Graf}}, \bibinfo {author}
  {\bibfnamefont {S.~M.~L.}\ \bibnamefont {Teicher}}, \bibinfo {author}
  {\bibfnamefont {R.}~\bibnamefont {Seshadri}}, \ and\ \bibinfo {author}
  {\bibfnamefont {S.~D.}\ \bibnamefont {Wilson}},\ }\bibfield  {title}
  {\enquote {\bibinfo {title} {Superconductivity in the ${\mathbb{z}}_{2}$
  kagome metal ${\mathrm{kv}}_{3}{\mathrm{sb}}_{5}$},}\ }\href {\doibase
  10.1103/PhysRevMaterials.5.034801} {\bibfield  {journal} {\bibinfo  {journal}
  {Phys. Rev. Materials}\ }\textbf {\bibinfo {volume} {5}},\ \bibinfo {pages}
  {034801} (\bibinfo {year} {2021})}\BibitemShut {NoStop}%
\bibitem [{\citenamefont {{Zhao}}\ \emph
  {et~al.}(2021{\natexlab{b}})\citenamefont {{Zhao}}, \citenamefont {{Wang}},
  \citenamefont {{Xia}}, \citenamefont {{Yin}}, \citenamefont {{Ni}},
  \citenamefont {{Huang}}, \citenamefont {{Tu}}, \citenamefont {{Tao}},
  \citenamefont {{Tu}}, \citenamefont {{Gong}}, \citenamefont {{Lei}},
  \citenamefont {{Guo}}, \citenamefont {{Yang}},\ and\ \citenamefont
  {{Li}}}]{zhao21axsc}%
  \BibitemOpen
  \bibfield  {author} {\bibinfo {author} {\bibfnamefont {C.~C.}\ \bibnamefont
  {{Zhao}}}, \bibinfo {author} {\bibfnamefont {L.~S.}\ \bibnamefont {{Wang}}},
  \bibinfo {author} {\bibfnamefont {W.}~\bibnamefont {{Xia}}}, \bibinfo
  {author} {\bibfnamefont {Q.~W.}\ \bibnamefont {{Yin}}}, \bibinfo {author}
  {\bibfnamefont {J.~M.}\ \bibnamefont {{Ni}}}, \bibinfo {author}
  {\bibfnamefont {Y.~Y.}\ \bibnamefont {{Huang}}}, \bibinfo {author}
  {\bibfnamefont {C.~P.}\ \bibnamefont {{Tu}}}, \bibinfo {author}
  {\bibfnamefont {Z.~C.}\ \bibnamefont {{Tao}}}, \bibinfo {author}
  {\bibfnamefont {Z.~J.}\ \bibnamefont {{Tu}}}, \bibinfo {author}
  {\bibfnamefont {C.~S.}\ \bibnamefont {{Gong}}}, \bibinfo {author}
  {\bibfnamefont {H.~C.}\ \bibnamefont {{Lei}}}, \bibinfo {author}
  {\bibfnamefont {Y.~F.}\ \bibnamefont {{Guo}}}, \bibinfo {author}
  {\bibfnamefont {X.~F.}\ \bibnamefont {{Yang}}}, \ and\ \bibinfo {author}
  {\bibfnamefont {S.~Y.}\ \bibnamefont {{Li}}},\ }\bibfield  {title} {\enquote
  {\bibinfo {title} {{Nodal superconductivity and superconducting domes in the
  topological Kagome metal CsV3Sb5}},}\ }\href@noop {} {\bibfield  {journal}
  {\bibinfo  {journal} {arXiv e-prints}\ ,\ \bibinfo {eid} {arXiv:2102.08356}}
  (\bibinfo {year} {2021}{\natexlab{b}})},\ \Eprint
  {http://arxiv.org/abs/2102.08356} {arXiv:2102.08356 [cond-mat.supr-con]}
  \BibitemShut {NoStop}%
\bibitem [{\citenamefont {Chen}\ \emph {et~al.}(2021)\citenamefont {Chen},
  \citenamefont {Wang}, \citenamefont {Yin}, \citenamefont {Gu}, \citenamefont
  {Jiang}, \citenamefont {Tu}, \citenamefont {Gong}, \citenamefont {Uwatoko},
  \citenamefont {Sun}, \citenamefont {Lei}, \citenamefont {Hu},\ and\
  \citenamefont {Cheng}}]{chen21prl}%
  \BibitemOpen
  \bibfield  {author} {\bibinfo {author} {\bibfnamefont {K.~Y.}\ \bibnamefont
  {Chen}}, \bibinfo {author} {\bibfnamefont {N.~N.}\ \bibnamefont {Wang}},
  \bibinfo {author} {\bibfnamefont {Q.~W.}\ \bibnamefont {Yin}}, \bibinfo
  {author} {\bibfnamefont {Y.~H.}\ \bibnamefont {Gu}}, \bibinfo {author}
  {\bibfnamefont {K.}~\bibnamefont {Jiang}}, \bibinfo {author} {\bibfnamefont
  {Z.~J.}\ \bibnamefont {Tu}}, \bibinfo {author} {\bibfnamefont {C.~S.}\
  \bibnamefont {Gong}}, \bibinfo {author} {\bibfnamefont {Y.}~\bibnamefont
  {Uwatoko}}, \bibinfo {author} {\bibfnamefont {J.~P.}\ \bibnamefont {Sun}},
  \bibinfo {author} {\bibfnamefont {H.~C.}\ \bibnamefont {Lei}}, \bibinfo
  {author} {\bibfnamefont {J.~P.}\ \bibnamefont {Hu}}, \ and\ \bibinfo {author}
  {\bibfnamefont {J.-G.}\ \bibnamefont {Cheng}},\ }\bibfield  {title} {\enquote
  {\bibinfo {title} {Double superconducting dome and triple enhancement of
  ${T}_{c}$ in the kagome superconductor ${\mathrm{csv}}_{3}{\mathrm{sb}}_{5}$
  under high pressure},}\ }\href {\doibase 10.1103/PhysRevLett.126.247001}
  {\bibfield  {journal} {\bibinfo  {journal} {Phys. Rev. Lett.}\ }\textbf
  {\bibinfo {volume} {126}},\ \bibinfo {pages} {247001} (\bibinfo {year}
  {2021})}\BibitemShut {NoStop}%
\bibitem [{\citenamefont {{Duan}}\ \emph {et~al.}(2021)\citenamefont {{Duan}},
  \citenamefont {{Nie}}, \citenamefont {{Luo}}, \citenamefont {{Yu}},
  \citenamefont {{Ortiz}}, \citenamefont {{Yin}}, \citenamefont {{Su}},
  \citenamefont {{Du}}, \citenamefont {{Wang}}, \citenamefont {{Chen}},
  \citenamefont {{Lu}}, \citenamefont {{Ying}}, \citenamefont {{Wilson}},
  \citenamefont {{Chen}}, \citenamefont {{Song}},\ and\ \citenamefont
  {{Yuan}}}]{duan21ax}%
  \BibitemOpen
  \bibfield  {author} {\bibinfo {author} {\bibfnamefont {W.}~\bibnamefont
  {{Duan}}}, \bibinfo {author} {\bibfnamefont {Z.}~\bibnamefont {{Nie}}},
  \bibinfo {author} {\bibfnamefont {S.}~\bibnamefont {{Luo}}}, \bibinfo
  {author} {\bibfnamefont {F.}~\bibnamefont {{Yu}}}, \bibinfo {author}
  {\bibfnamefont {B.~R.}\ \bibnamefont {{Ortiz}}}, \bibinfo {author}
  {\bibfnamefont {L.}~\bibnamefont {{Yin}}}, \bibinfo {author} {\bibfnamefont
  {H.}~\bibnamefont {{Su}}}, \bibinfo {author} {\bibfnamefont {F.}~\bibnamefont
  {{Du}}}, \bibinfo {author} {\bibfnamefont {A.}~\bibnamefont {{Wang}}},
  \bibinfo {author} {\bibfnamefont {Y.}~\bibnamefont {{Chen}}}, \bibinfo
  {author} {\bibfnamefont {X.}~\bibnamefont {{Lu}}}, \bibinfo {author}
  {\bibfnamefont {J.}~\bibnamefont {{Ying}}}, \bibinfo {author} {\bibfnamefont
  {S.~D.}\ \bibnamefont {{Wilson}}}, \bibinfo {author} {\bibfnamefont
  {X.}~\bibnamefont {{Chen}}}, \bibinfo {author} {\bibfnamefont
  {Y.}~\bibnamefont {{Song}}}, \ and\ \bibinfo {author} {\bibfnamefont
  {H.}~\bibnamefont {{Yuan}}},\ }\bibfield  {title} {\enquote {\bibinfo {title}
  {{Nodeless superconductivity in the kagome metal CsV$_3$Sb$_5$}},}\
  }\href@noop {} {\bibfield  {journal} {\bibinfo  {journal} {arXiv e-prints}\
  ,\ \bibinfo {eid} {arXiv:2103.11796}} (\bibinfo {year} {2021})},\ \Eprint
  {http://arxiv.org/abs/2103.11796} {arXiv:2103.11796 [cond-mat.supr-con]}
  \BibitemShut {NoStop}%
\bibitem [{\citenamefont {Zhang}\ \emph {et~al.}(2021)\citenamefont {Zhang},
  \citenamefont {Chen}, \citenamefont {Zhou}, \citenamefont {Yuan},
  \citenamefont {Wang}, \citenamefont {Wang}, \citenamefont {Yang},
  \citenamefont {An}, \citenamefont {Zhang}, \citenamefont {Zhu}, \citenamefont
  {Zhou}, \citenamefont {Chen}, \citenamefont {Zhou},\ and\ \citenamefont
  {Yang}}]{zhang21prb}%
  \BibitemOpen
  \bibfield  {author} {\bibinfo {author} {\bibfnamefont {Z.}~\bibnamefont
  {Zhang}}, \bibinfo {author} {\bibfnamefont {Z.}~\bibnamefont {Chen}},
  \bibinfo {author} {\bibfnamefont {Y.}~\bibnamefont {Zhou}}, \bibinfo {author}
  {\bibfnamefont {Y.}~\bibnamefont {Yuan}}, \bibinfo {author} {\bibfnamefont
  {S.}~\bibnamefont {Wang}}, \bibinfo {author} {\bibfnamefont {J.}~\bibnamefont
  {Wang}}, \bibinfo {author} {\bibfnamefont {H.}~\bibnamefont {Yang}}, \bibinfo
  {author} {\bibfnamefont {C.}~\bibnamefont {An}}, \bibinfo {author}
  {\bibfnamefont {L.}~\bibnamefont {Zhang}}, \bibinfo {author} {\bibfnamefont
  {X.}~\bibnamefont {Zhu}}, \bibinfo {author} {\bibfnamefont {Y.}~\bibnamefont
  {Zhou}}, \bibinfo {author} {\bibfnamefont {X.}~\bibnamefont {Chen}}, \bibinfo
  {author} {\bibfnamefont {J.}~\bibnamefont {Zhou}}, \ and\ \bibinfo {author}
  {\bibfnamefont {Z.}~\bibnamefont {Yang}},\ }\bibfield  {title} {\enquote
  {\bibinfo {title} {Pressure-induced reemergence of superconductivity in the
  topological kagome metal $\mathrm{Cs}{\mathrm{v}}_{3}{\mathrm{sb}}_{5}$},}\
  }\href {\doibase 10.1103/PhysRevB.103.224513} {\bibfield  {journal} {\bibinfo
   {journal} {Phys. Rev. B}\ }\textbf {\bibinfo {volume} {103}},\ \bibinfo
  {pages} {224513} (\bibinfo {year} {2021})}\BibitemShut {NoStop}%
\bibitem [{\citenamefont {Sun}\ \emph {et~al.}(2009)\citenamefont {Sun},
  \citenamefont {Yao}, \citenamefont {Fradkin},\ and\ \citenamefont
  {Kivelson}}]{sun09prl}%
  \BibitemOpen
  \bibfield  {author} {\bibinfo {author} {\bibfnamefont {K.}~\bibnamefont
  {Sun}}, \bibinfo {author} {\bibfnamefont {H.}~\bibnamefont {Yao}}, \bibinfo
  {author} {\bibfnamefont {E.}~\bibnamefont {Fradkin}}, \ and\ \bibinfo
  {author} {\bibfnamefont {S.~A.}\ \bibnamefont {Kivelson}},\ }\bibfield
  {title} {\enquote {\bibinfo {title} {Topological insulators and nematic
  phases from spontaneous symmetry breaking in 2d fermi systems with a
  quadratic band crossing},}\ }\href {\doibase 10.1103/PhysRevLett.103.046811}
  {\bibfield  {journal} {\bibinfo  {journal} {Phys. Rev. Lett.}\ }\textbf
  {\bibinfo {volume} {103}},\ \bibinfo {pages} {046811} (\bibinfo {year}
  {2009})}\BibitemShut {NoStop}%
\bibitem [{\citenamefont {Chern}\ and\ \citenamefont
  {Batista}(2012)}]{chern12prl}%
  \BibitemOpen
  \bibfield  {author} {\bibinfo {author} {\bibfnamefont {G.-W.}\ \bibnamefont
  {Chern}}\ and\ \bibinfo {author} {\bibfnamefont {C.~D.}\ \bibnamefont
  {Batista}},\ }\bibfield  {title} {\enquote {\bibinfo {title} {Spontaneous
  quantum hall effect via a thermally induced quadratic fermi point},}\ }\href
  {\doibase 10.1103/PhysRevLett.109.156801} {\bibfield  {journal} {\bibinfo
  {journal} {Phys. Rev. Lett.}\ }\textbf {\bibinfo {volume} {109}},\ \bibinfo
  {pages} {156801} (\bibinfo {year} {2012})}\BibitemShut {NoStop}%
\bibitem [{\citenamefont {Kiesel}\ and\ \citenamefont
  {Thomale}(2012)}]{kiesel12prb}%
  \BibitemOpen
  \bibfield  {author} {\bibinfo {author} {\bibfnamefont {M.~L.}\ \bibnamefont
  {Kiesel}}\ and\ \bibinfo {author} {\bibfnamefont {R.}~\bibnamefont
  {Thomale}},\ }\bibfield  {title} {\enquote {\bibinfo {title} {Sublattice
  interference in the kagome hubbard model},}\ }\href {\doibase
  10.1103/PhysRevB.86.121105} {\bibfield  {journal} {\bibinfo  {journal} {Phys.
  Rev. B}\ }\textbf {\bibinfo {volume} {86}},\ \bibinfo {pages} {121105}
  (\bibinfo {year} {2012})}\BibitemShut {NoStop}%
\bibitem [{\citenamefont {Haldane}(1988)}]{haldane88prl}%
  \BibitemOpen
  \bibfield  {author} {\bibinfo {author} {\bibfnamefont {F.~D.~M.}\
  \bibnamefont {Haldane}},\ }\bibfield  {title} {\enquote {\bibinfo {title}
  {Model for a quantum hall effect without landau levels: Condensed-matter
  realization of the "parity anomaly"},}\ }\href {\doibase
  10.1103/PhysRevLett.61.2015} {\bibfield  {journal} {\bibinfo  {journal}
  {Phys. Rev. Lett.}\ }\textbf {\bibinfo {volume} {61}},\ \bibinfo {pages}
  {2015} (\bibinfo {year} {1988})}\BibitemShut {NoStop}%
\bibitem [{\citenamefont {Hasan}\ and\ \citenamefont
  {Kane}(2010)}]{hasan10rmp}%
  \BibitemOpen
  \bibfield  {author} {\bibinfo {author} {\bibfnamefont {M.~Z.}\ \bibnamefont
  {Hasan}}\ and\ \bibinfo {author} {\bibfnamefont {C.~L.}\ \bibnamefont
  {Kane}},\ }\bibfield  {title} {\enquote {\bibinfo {title} {Colloquium:
  Topological insulators},}\ }\href {\doibase 10.1103/RevModPhys.82.3045}
  {\bibfield  {journal} {\bibinfo  {journal} {Rev. Mod. Phys.}\ }\textbf
  {\bibinfo {volume} {82}},\ \bibinfo {pages} {3045} (\bibinfo {year}
  {2010})}\BibitemShut {NoStop}%
\bibitem [{\citenamefont {Fukui}\ \emph {et~al.}(2005)\citenamefont {Fukui},
  \citenamefont {Hatsugai},\ and\ \citenamefont {Suzuki}}]{fukui05jpsp}%
  \BibitemOpen
  \bibfield  {author} {\bibinfo {author} {\bibfnamefont {T.}~\bibnamefont
  {Fukui}}, \bibinfo {author} {\bibfnamefont {Y.}~\bibnamefont {Hatsugai}}, \
  and\ \bibinfo {author} {\bibfnamefont {H.}~\bibnamefont {Suzuki}},\
  }\bibfield  {title} {\enquote {\bibinfo {title} {Chern numbers in discretized
  brillouin zone: Efficient method of computing (spin) hall conductances},}\
  }\href {\doibase 10.1143/JPSJ.74.1674} {\bibfield  {journal} {\bibinfo
  {journal} {J. Phys. Soc. Jpn.}\ }\textbf {\bibinfo {volume} {74}},\ \bibinfo
  {pages} {1674} (\bibinfo {year} {2005})}\BibitemShut {NoStop}%
\bibitem [{\citenamefont {Cooper}\ \emph {et~al.}(2019)\citenamefont {Cooper},
  \citenamefont {Dalibard},\ and\ \citenamefont {Spielman}}]{cooper19rmp}%
  \BibitemOpen
  \bibfield  {author} {\bibinfo {author} {\bibfnamefont {N.~R.}\ \bibnamefont
  {Cooper}}, \bibinfo {author} {\bibfnamefont {J.}~\bibnamefont {Dalibard}}, \
  and\ \bibinfo {author} {\bibfnamefont {I.~B.}\ \bibnamefont {Spielman}},\
  }\bibfield  {title} {\enquote {\bibinfo {title} {Topological bands for
  ultracold atoms},}\ }\href {\doibase 10.1103/RevModPhys.91.015005} {\bibfield
   {journal} {\bibinfo  {journal} {Rev. Mod. Phys.}\ }\textbf {\bibinfo
  {volume} {91}},\ \bibinfo {pages} {015005} (\bibinfo {year}
  {2019})}\BibitemShut {NoStop}%
\bibitem [{\citenamefont {{Lin}}(2021)}]{lin21axhoti}%
  \BibitemOpen
  \bibfield  {author} {\bibinfo {author} {\bibfnamefont {Y.-P.}\ \bibnamefont
  {{Lin}}},\ }\bibfield  {title} {\enquote {\bibinfo {title} {{Higher-order
  topological insulators from $3Q$ charge bond orders on hexagonal lattices: A
  hint to kagome metals}},}\ }\href@noop {} {\bibfield  {journal} {\bibinfo
  {journal} {arXiv e-prints}\ ,\ \bibinfo {eid} {arXiv:2106.09717}} (\bibinfo
  {year} {2021})},\ \Eprint {http://arxiv.org/abs/2106.09717} {arXiv:2106.09717
  [cond-mat.str-el]} \BibitemShut {NoStop}%
\bibitem [{\citenamefont {{Feng}}\ \emph {et~al.}(2021)\citenamefont {{Feng}},
  \citenamefont {{Jiang}}, \citenamefont {{Wang}},\ and\ \citenamefont
  {{Hu}}}]{feng21ax}%
  \BibitemOpen
  \bibfield  {author} {\bibinfo {author} {\bibfnamefont {X.}~\bibnamefont
  {{Feng}}}, \bibinfo {author} {\bibfnamefont {K.}~\bibnamefont {{Jiang}}},
  \bibinfo {author} {\bibfnamefont {Z.}~\bibnamefont {{Wang}}}, \ and\ \bibinfo
  {author} {\bibfnamefont {J.}~\bibnamefont {{Hu}}},\ }\bibfield  {title}
  {\enquote {\bibinfo {title} {{Chiral flux phase in the Kagome superconductor
  AV$_3$Sb$_5$}},}\ }\href@noop {} {\bibfield  {journal} {\bibinfo  {journal}
  {arXiv e-prints}\ ,\ \bibinfo {eid} {arXiv:2103.07097}} (\bibinfo {year}
  {2021})},\ \Eprint {http://arxiv.org/abs/2103.07097} {arXiv:2103.07097
  [cond-mat.supr-con]} \BibitemShut {NoStop}%
\bibitem [{\citenamefont {{Denner}}\ \emph {et~al.}(2021)\citenamefont
  {{Denner}}, \citenamefont {{Thomale}},\ and\ \citenamefont
  {{Neupert}}}]{denner21ax}%
  \BibitemOpen
  \bibfield  {author} {\bibinfo {author} {\bibfnamefont {M.~M.}\ \bibnamefont
  {{Denner}}}, \bibinfo {author} {\bibfnamefont {R.}~\bibnamefont {{Thomale}}},
  \ and\ \bibinfo {author} {\bibfnamefont {T.}~\bibnamefont {{Neupert}}},\
  }\bibfield  {title} {\enquote {\bibinfo {title} {{Analysis of charge order in
  the kagome metal $A$V$_3$Sb$_5$ ($A=$K,Rb,Cs)}},}\ }\href@noop {} {\bibfield
  {journal} {\bibinfo  {journal} {arXiv e-prints}\ ,\ \bibinfo {eid}
  {arXiv:2103.14045}} (\bibinfo {year} {2021})},\ \Eprint
  {http://arxiv.org/abs/2103.14045} {arXiv:2103.14045 [cond-mat.str-el]}
  \BibitemShut {NoStop}%
\end{thebibliography}%

\end{document}